\documentclass[a4paper,12pt]{article}

\usepackage[T1]{fontenc}

\usepackage[a4paper,margin=1in]{geometry}

\usepackage{palatino}

\usepackage{amsmath,amssymb,mathtools,amsthm,bbm}

\usepackage{newpxmath}

\usepackage{graphicx}
\usepackage{subcaption}
\usepackage{booktabs}
\usepackage{threeparttable}
\usepackage{makecell}
\usepackage{adjustbox}
\usepackage{float}
\usepackage{placeins}
\usepackage{pdflscape}
\usepackage{multirow}

\usepackage[dvipsnames,table]{xcolor}
\definecolor{darkblue}{HTML}{004080}
\definecolor{cornellred}{rgb}{0.7,0.11,0.11}

\usepackage{enumitem}
\usepackage{indentfirst}
\usepackage{setspace}
\usepackage{abstract}
\usepackage{epigraph}
\usepackage{fancyhdr}
\usepackage{dirtytalk}
\usepackage{array}

\usepackage[sort]{natbib}
\usepackage[nottoc]{tocbibind}
\usepackage[titletoc]{appendix}

\usepackage{tikz}
\usepackage{tikz-cd}

\usepackage{listings}
\usepackage{fancyvrb}

\usepackage{hyperref}
\hypersetup{
    colorlinks=true,
    breaklinks=true,
    linkcolor=darkblue,
    urlcolor=darkblue,
    anchorcolor=darkblue,
    citecolor=darkblue
}
\usepackage{cleveref}

\newcommand{\sym}[1]{\ifmmode^{#1}\else\(^{#1}\)\fi}

\newcounter{thm}

\newcounter{hyp}[section]

\title{
    \vspace{-1.5cm}
    \LARGE Blameocracy:\\
    Causal Rhetoric in Politics\thanks{
        \noindent Francesco Bilotta: Bocconi University, francesco.bilotta2@phd.unibocconi.it;
        Alberto Binetti: Princeton University, abinetti@princeton.edu;
        Giacomo Manferdini: Bocconi University; giacomo.manferdini@phd.unibocconi.it.
        Special thanks go to Massimo Morelli, Nicola Gennaioli, Egon Tripodi, and Carlo Schwarz for their guidance, advice, and support throughout the project. We are grateful to Enrico D. Turri for his input at earlier stages. We also thank Luca Braghieri, Felix Chopra, Sarah Eichmeyer, Claudio Ferraz, Gloria Gennaro, Dirk Hovy, Rafael Jiménez-Durán, Tania Lombrozo, Jaime Marques Pereira, Edoardo Teso, and Ekaterina Zhuravskaya for their helpful comments and suggestions, as well as participants at the Munich Workshop on Beliefs Narratives and Memory, 2nd Bocconi-CCA-Cornell Political Economy Workshop, Princeton Junior PE Seminar, Workshop on Text-as-Data in Economics at Lancaster, 2nd Verona Early Career Workshop in Economics, Potsdam Text-as-Data in Behavioral Economics Workshop, XIV Alghero IBEO Workshop, 1st Berlin Micro Theory and Behavioral Economics PhD Conference, 10th Monash-Paris-Warwick-Zurich-CEPR Text-as-Data Workshop, XXIII RIDGE-LACEA Political Economy Workshop, 9th Economics of Media Bias Workshop, 2nd Berlin PhD Conference in Economics, and Cognitive Economics Workshop.
    }
}

\author{
    \centering
    \begin{tabular}{@{}>{\centering\arraybackslash}p{0.3\textwidth}>{\centering\arraybackslash}p{0.3\textwidth}>{\centering\arraybackslash}p{0.3\textwidth}@{}}
        Francesco Bilotta & Alberto Binetti & Giacomo Manferdini \\[-0.15cm]
        & & \textit{Job Market Paper}
    \end{tabular}%
}

\date{
    \vspace{0.25cm}
    \today\\[0.1cm]
    \href{https://manferdinig.github.io/Market/Manferdini_JMP_Blameocracy.pdf}{Click here for the latest version}
}

\begin{document}
\maketitle
\thispagestyle{empty}

\vspace{-0.25cm}
\begin{abstract}
\vspace{-0.75cm}
\noindent 

Democratic politics requires voters to infer how political actions map into outcomes, yet these causal links are often unclear, giving politicians scope to shape voters' causal accounts. This paper studies the supply and returns of causal rhetoric in politics: politicians' language linking political actions to outcomes through blame and credit. We fine-tune a language model to measure its use in 4.2 million congressional tweets spanning 2012--2023. Causal rhetoric is central to politicians' communication: it accounts for more than one third of tweets, is the language of political disagreement, and rises over time. The political returns of causal rhetoric are concentrated in blame. Combining campaign contributions with quasi-exogenous geographic variation in Twitter adoption, we find that exposure to blame has the largest effects on small donations. Using voting intentions from survey data and an event-study design around blame-only and credit-only days, we find that politicians' support rises after blame-only days. Finally, causal rhetoric appears rarely supported by economic knowledge, particularly blame.

\end{abstract}

\newpage
\pagenumbering{arabic}


\begin{quote}
    \footnotesize
    \flushright 
    ``To err is human. To blame someone else is politics.''\\
    \hfill Hubert H. Humphrey, Vice President of the United States (1965--1969)
\end{quote}

\section{Introduction}

Democratic politics asks voters to make a difficult causal judgment. They must form beliefs about how political actions map into outcomes: who is responsible for what has happened, and what consequences political actors are likely to produce in the future \citep{schmitter1991democracy}. Yet this mapping is rarely transparent. Policies operate in complex environments, outcomes have many causes, and responsibility is often shared across parties, institutions, and levels of government. Voters therefore face the problem of reconstructing a mapping between political actions and outcomes that is politically consequential but hard to infer \citep{achen2016democracy}.

This difficulty creates a rhetorical margin: politicians can shape voters' understanding of this mapping. We call this margin \textit{causal rhetoric}: language that links the actual or potential action of a political actor to an outcome. When the outcome is negative, politicians assign \textit{blame}; when it is positive, they claim \textit{credit}. In both cases, causal rhetoric supplies voters with an account of how political actions and outcomes are connected, bringing a well-documented persuasive force to a central problem of democratic politics. This force comes from the way causal accounts shape the information people seek, remember, and act on \citep{graeber2026explanations,huning2022using,andre2026narratives,bursztyn2023opinions}.\footnote{This evidence is consistent with a broader literature in cognitive psychology emphasizing that causal thinking is central to human reasoning \citep[e.g.,][]{sloman2015causality,lombrozo2017causal,chater2016under}.} In politics, these accounts are a first step in the accountability process: they tell voters which outcomes to link to which political actions. Yet we know little about how politicians supply causal rhetoric or the political returns they gain from using it.

This paper studies these questions in the context of U.S. congressional communication on Twitter. We fine-tune a language model to build a content-based measure of causal rhetoric, classifying tweets by whether they assign blame or claim credit. We then combine this measure with data on campaign contributions and voting intentions to document the supply of causal rhetoric and estimate its political returns. The results show that causal rhetoric is a central part of politicians' communication and that its political returns are concentrated in blame. Finally, we also assess the economic grounding of these claims and find that they are rarely supported by economic knowledge, especially when they assign blame.

The measurement starts from a corpus of 4.2 million tweets posted by Democratic and Republican members of Congress between 2012 and 2023. We hand-label a training set of tweets and use these labels to fine-tune a RoBERTa-large model pre-trained on Twitter data \citep{loureiro2022timelms}, classifying each tweet as blame, credit, or neither. The resulting measure performs well against hand-coded validation data and aligns closely with independent external classifications. Additional validation shows that blame and credit exhibit the linguistic patterns expected of causal attribution and are not simply a relabeling of previously studied dimensions of political language. This yields a scalable measure of causal rhetoric in politicians' communication on social media.

We first establish that causal rhetoric is a central part of politicians' communication on social media. More than one third of tweets contain causal rhetoric: 16 percent assign blame and 20 percent claim credit. Its use is more concentrated in policy domains than in sociocultural ones, consistent with politicians using it where political actions can be more naturally linked to outcomes. Causal rhetoric is also the language of political disagreement. The sentiment gap between the ruling party and the opposition is much larger in tweets that assign blame or claim credit than in other tweets, indicating that politicians' evaluative disagreement is especially concentrated in causal claims. Finally, the aggregate share of blame and credit rises over time, from roughly one fifth of congressional tweets at the beginning of the sample to nearly one half by the end, with both components increasing.

This prevalence motivates the paper's main returns question: does causal rhetoric generate political support? Returns reveal whether this widespread rhetorical input is politically rewarded, rather than merely supplied, and whether it affects citizens' political behavior. We find that these returns are concentrated in blame. Exposure to blame generates sizable and consistent gains in support, while credit produces no comparable aggregate response. We establish this pattern using two complementary sources of evidence: small campaign contributions and survey data on voting intentions.

Our first source of evidence is campaign contributions. We link our rhetoric measure to more than 40 million small donations, defined as individual contributions below \$1,000 \citep{bonica2024dime,petrova2021social,boken2026returns}. To estimate returns, we use a cross-county shift-share-style exposure design \citep{goldsmith2020bartik,borusyak2025practical} that combines politician-month variation in blame and credit with quasi-exogenous geographic variation in Twitter adoption induced by the platform's early diffusion after the 2007 South by Southwest festival \citep{muller2023hashtag,fujiwara2024effect}. The identifying comparison asks whether counties with greater quasi-exogenous Twitter exposure donate more to the same politician in months when that politician uses more causal rhetoric, net of politician-month, politician-county, and county-month fixed effects. Because the design relies on quasi-exogenous differences in exposure rather than on short-run timing around tweets, it does not require politicians to choose blame or credit at random.

The donation estimates show asymmetric returns, larger for blame. A one-standard-deviation increase in the monthly share of blame tweets raises small-donation revenue by roughly 2.6 percent in a county with average Twitter adoption, while the corresponding effect of credit is about 0.5 percent. These aggregate effects operate through different margins of support. Blame expands the donor base, increasing the number of donors by about 1.6 percent, whereas credit has essentially no effect on donor counts and its small return is concentrated in higher average giving. The returns also differ by donor ideology, suggesting that blame and credit activate different constituencies within the donor base: ideologically extreme donors respond positively to blame and negatively to credit, while moderate donors display the opposite pattern.

We complement the donation evidence with survey data on voting intentions in the 2020 House elections to study political support in a broader population beyond campaign donors. Using 92,211 responses from Nationscape \citep{tausanovitch2021nationscape}, we link respondents to the Twitter communication of the sitting representative in their congressional district. This setting lets us measure political support directly as intended vote choice and use a short-run timing comparison. We implement an event-study design around days when a politician's Twitter communication consists entirely of one rhetorical style. The design asks whether support for the same politician changes after blame-only or credit-only days, net of district-month, date, partisan-alignment, and respondent-demographic fixed effects.

The survey estimates show an asymmetric response in intended vote choice. Support for the politician rises by 1 percentage point in the week after blame-only days. The corresponding estimate after credit-only days is smaller and imprecisely estimated. A dynamic version of the design shows flat pre-trends before blame-only days, followed by a short-run increase in support that fades within two weeks. The response is concentrated among respondents who report using social media for political news: among these respondents, support rises by 1.4 percentage points after blame-only days, compared with a smaller and imprecise estimate among other respondents. The survey evidence therefore points to the same asymmetry as the donation evidence and reinforces the social-media exposure channel.

We find no evidence that the returns to blame simply reflect exposure to negativity. If negativity alone drove these returns, we would expect non-causal negative rhetoric, which features a negative tone but does not link political actions to outcomes, to generate positive political returns as well. To assess this possibility, we repeat both empirical analyses using non-causal negative and positive rhetoric in place of blame and credit. Across campaign contributions and voting intentions, non-causal negative rhetoric does not reproduce the statistically significant positive returns observed for blame.

Having established the supply and political returns of causal rhetoric, we ask a further question about the accounts politicians supply: how well grounded are the causal claims they are trying to convince voters of? We study this question in the economic domain, where causal rhetoric is common and where claims linking political actions to outcomes can be assessed against standard macroeconomic reasoning. Using an AI-assisted coding procedure, we assess whether economic causal claims are supported as stated or instead require qualifications that the tweet does not provide. Only about 10 percent are supported as stated. Roughly 90 percent require qualification, while outright unsupported claims are rare. The need for qualification is substantially greater for blame than for credit.

Before concluding, we also document two additional patterns in how causal rhetoric is transmitted and deployed. Blame receives substantially greater amplification on social media, attracting nearly four times as many retweets as other rhetorical styles, a pattern consistent with its larger political returns. Causal rhetoric is also deployed in politically convenient ways: economic blame becomes more common when the opposing party is legislatively active, while credit becomes more common when one's own party is active.

Taken together, the evidence shows that causal rhetoric is a major input into democratic politics. Politicians widely supply voters with accounts of how political actions map into outcomes, and these accounts generate political returns when they assign blame. At the same time, causal rhetoric, especially blame, is rarely supported by economic knowledge. Blame therefore stands out across these findings: it generates the largest political returns while being the least economically grounded.

\paragraph*{Related Literature}

We contribute to the literature on accountability in political economy by studying how politicians seek to shape the attributions on which accountability rests. A large body of work studies how accountability shapes politicians' behavior, showing, for example, how term limits and press coverage discipline elected officials \citep{besley1995does,besley2006principled,ferraz2011electoral,snyder2010press,ashworth2012electoral,maskin2004politician,canes2001leadership,alonso2026ambiguous}. A related strand studies how voters use realized outcomes to evaluate incumbents, emphasizing both the logic and the limits of retrospective voting \citep{healy2013retrospective,ferraz2008exposing,achen2016democracy}. We study an upstream margin in this process: politicians' use of rhetoric that links political actions to outcomes. We document the prevalence of this rhetoric and estimate its political returns.

This paper adds to the political-economy literature that studies the effects of social media on political behavior by identifying causal rhetoric as a specific content channel through which online communication affects political support. A growing body of work shows that social media affects political support and political behavior \citep{fujiwara2024effect,muller2023hashtag,zhuravskaya2020political,rotesi2019impact}. Closest to our paper is work on Twitter and campaign finance: \citet{petrova2021social} show that politicians benefit from adopting Twitter, especially entrants and candidates in high-adoption states, while \citet{boken2026returns} document a fundraising premium when politicians go viral on Twitter. We shift attention from platform adoption, diffusion, and virality to the content of politicians' messages. Rather than treating social-media exposure as a single reduced-form treatment, we identify blame and credit as rhetorical inputs through which politicians' online communication generates political support.

We also contribute to the literature in political economy that measures narratives in observational data. Existing work has mainly focused on narratives surrounding specific topics, such as climate change or stock-market crashes \citep{goetzmann2024crash,gehring2025virality,macaulay2024narrative,bilotta2026policy}. We take a more agnostic approach, measuring a broad reduced-form feature of political narratives: whether politicians connect political actions to outcomes, regardless of the topic under discussion. We then link these narratives to political outcomes outside social media, estimating their effects on campaign contributions and voting intentions. This measure connects to \citet{bilotta2026coarse}, where politicians compete by shifting blame and claiming credit when voters cannot fully recover the relationship between policies and outcomes. Our evidence documents both the prevalence of this rhetoric and the political returns it generates.

Finally, we contribute to the literature on political rhetoric by isolating the causal margin of political language: whether politicians link political actions to outcomes. A long-standing literature on negative campaigning studies whether campaigns attack opponents or promote the candidate \citep{lau2009negative}. We focus instead on blame as a specific form of negative rhetoric: language that connects political actors or their actions to bad outcomes. This excludes other negative content, such as personal attacks, insults, or moral condemnation, when it does not contain a causal claim. We also complement recent work on expressive dimensions of political language, such as emotionality, emotions, and moral values \citep{gennaro2022emotion,enke2020moral,algan2026emotions}. We instead identify causal rhetoric as a distinct dimension of political discourse.

\medskip
The rest of the paper proceeds as follows. \Cref{sec:data} describes the data. \Cref{sec:measurement} defines causal rhetoric through blame and credit, explains how we measure it, and validates the resulting classification. \Cref{sec:supply} documents the prevalence, political-disagreement patterns, and trends of causal rhetoric in congressional communication. \Cref{sec:returns} estimates its political returns using campaign-contribution and voting-intention data. \Cref{sec:content} assesses the economic grounding of causal rhetoric. \Cref{sec:additional} documents additional patterns in the amplification and politically convenient deployment of causal rhetoric. Finally, \Cref{sec:conclusion} concludes.
\section{Data}\label{sec:data}

Our analysis combines several datasets on political communication, campaign finance, and survey support.

\paragraph*{Tweets}

Our main dataset contains approximately 4.2 million tweets posted by Democratic and Republican members of the U.S. Congress between 2012 and 2023. We combine data from the \citeauthor{congresstweets} project and \citet{bellodi2025shift}. The resulting dataset covers the universe of tweets posted by Representatives from January 3, 2013, to July 11, 2023, and by Senators from June 21, 2017, to July 11, 2023. It also covers all tweets posted over the full 2012 calendar year by Representatives running for re-election that year. The final sample includes 4,198,455 tweets from 900 unique politicians and 1,789 unique Twitter accounts.\footnote{We exclude retweets because they do not represent original content. We retain quoted tweets because they also contain user-generated text.} We further enrich these data with demographic and ideological information from VoteView and Wikipedia, including age, gender, race, education, and NOMINATE scores. \Cref{tab:descriptives} reports descriptive statistics at both the tweet and politician levels. Republicans account for a slight majority of politicians in the sample, but most tweets are posted by Democrats, who---as also documented by \citet{fujiwara2024effect}---are more active on Twitter. Most politicians in the dataset hold at least a bachelor's degree, the average age is just under 60, and about one quarter are women.

\paragraph*{Donations}

We collect campaign contribution data from the Database on Ideology, Money in Politics, and Elections compiled by \citet{bonica2024dime}. This dataset includes more than 850 million donations made by individuals and organizations to candidates in local, state, and federal elections between 1979 and 2024. We restrict the sample to individual contributions directed to candidates for the U.S. House of Representatives and U.S. Senate in the 2012--2024 election cycles. Following the literature, we define small donations as contributions below \$1,000 \citep{petrova2021social,boken2026returns}. In the analysis, we link these donations to politicians in our Twitter data, so the effective sample period is determined by the overlap between donation records and observed tweeting behavior and therefore runs through 2023. The resulting matched sample contains approximately 40.1 million donations from 4.58 million donors to 733 politicians. \Cref{tab:descriptives_donations} reports descriptive statistics at both the donation and donor levels. The average donation is \$36.40, the median is \$15, and 90 percent of all contributions are below \$95. Over the sample period, each donor contributes \$318.62 on average, while the median donor contributes \$53.93.

\paragraph*{Voting Intentions}

We use Nationscape survey data for the 2020 House elections. Nationscape is a repeated cross-section that records respondents' congressional district, intended House vote choice, party identification, demographic characteristics, and recent political-news sources \citep{tausanovitch2021nationscape}. To ensure that the relevant general-election choice is defined, we restrict the sample to respondents interviewed after the relevant congressional primary in their state, using official primary dates from \citeauthor{fec2020primarydates}.\footnote{In states where a runoff can determine the nominee, we use the runoff date instead. Louisiana is treated separately because it does not hold a separate congressional primary; for Louisiana, we use the congressional ballot-access filing deadline.} We keep respondents interviewed through October 31, 2020, just before the general election, and restrict attention to districts in which the sitting representative is running for re-election, since our tweet data cover sitting legislators. This leaves us with 92,211 survey responses.

\section{Measurement}\label{sec:measurement}

The empirical analysis requires a tweet-level measure of causal rhetoric. We define causal rhetoric as language that links the actual or potential action of a political actor to an outcome. We then distinguish whether this link is framed negatively, as blame, or positively, as credit. This definition leads to a two-step measurement problem: identifying whether a tweet contains a causal claim, and classifying the tone of that claim.

We implement this measure at scale by fine-tuning a RoBERTa-large model. We then validate the resulting classifications in three ways. First, we show that the classifier performs well against hand-coded labels. Second, we compare our measure with external benchmarks from America's Political Pulse. Third, we document internal linguistic patterns consistent with the interpretation of blame and credit as distinct forms of causal rhetoric.

\subsection{Definition}

Our classification rests on two dimensions: causality and tone. A tweet is causal if it links the actual or potential action of a political actor to an outcome. Tone captures whether the tweet expresses a negative, neutral, or positive stance toward its subject.

Causality is binary. Political actors include politicians, political institutions, and politically aligned organizations. They exclude natural events, such as pandemics or disasters, and nonpolitical actors, such as scientific teams. A tweet is causal only if it links a politically meaningful action of such an actor to an outcome. The action may be explicit, as when the tweet refers to a particular bill, reform, or decision. It may also be implicit, as when the tweet attributes an outcome to a political actor while leaving the specific action unstated. In either case, the action may already have occurred, may be ongoing, or may be one the actor would take if in power. Outcomes may be economic, such as income or inflation, or social, such as civil liberties.

Tone captures the stance of the tweet toward its subject. We classify tone as negative, neutral, or positive. Negative and positive tone reflect unfavorable and favorable language, respectively. Neutral tone captures descriptive statements without evaluative content. Our manual annotation is designed to capture nuances that dictionary-based methods often miss.

Combining causality and tone yields our measure of rhetorical style. Tweets that are both causal and negative are coded as blame. Tweets that are both causal and positive are coded as credit. All remaining tweets, including noncausal tweets and causal tweets with neutral tone, are coded as none. \Cref{tab:sample_tweets_labeled} reports labeled examples.

\subsection{Classification}

To implement our measure of rhetorical style at scale, we begin by constructing a labeled dataset of 3,958 tweets, each annotated along the two dimensions introduced above: causality and tone. Because blame and credit tweets are likely to represent only a minority of the full corpus, a purely random sample would yield too few examples of the classes most relevant for training. This matters especially in our setting because fine-tuning performance depends on exposing the model to a sufficiently balanced set of examples across classes. To improve learning, we therefore combine random sampling with targeted oversampling, a common strategy in the classification literature \citep[e.g.,][]{he2009learning, talat2016hateful, davidson2017automated}. Specifically, we first generate benchmark examples of blame and credit tweets for Democrats and Republicans using ChatGPT-4o, solely to guide sampling. We then compute cosine similarity between these benchmark examples and the full tweet corpus using SBERT-mini embeddings \citep{reimers2019sentence}. Half of the tweets selected for annotation are drawn from those most similar to the benchmarks, balancing by party and rhetorical style, while the remaining half are sampled at random.

To assess annotation reliability, we select a subsample of 530 tweets and have each of them independently labeled by all three coders. For this jointly annotated subsample, labels for causality and tone are assigned by majority vote, with ties broken randomly. The resulting protocol delivers strong inter-annotator agreement on the rhetorical style measure: the average pairwise correlation is 0.73 and Fleiss' kappa is 0.64, typically interpreted as substantial agreement.\footnote{Fleiss' kappa is a standard statistic for assessing agreement among more than two coders when observations are assigned to mutually exclusive categories. It is the natural multi-coder extension of Cohen's kappa. Unlike simple percent agreement, it adjusts for the amount of agreement that would arise mechanically by chance, and therefore provides a more informative measure of annotation reliability.} Coders agree on rhetorical style in 67 percent of tweets, roughly six times the rate expected by chance.
 
We then fine-tune a RoBERTa-large model pre-trained on 154 million tweets \citep{loureiro2022timelms}.\footnote{Fine-tuning adapts a pre-trained language model to a specific downstream task by updating its parameters on a smaller labeled dataset. In our setting, this allows the model to learn how causal attributions and tone are expressed in political language.} We split the labeled corpus into 80 percent training data and 20 percent validation data.\footnote{We train for 10 epochs and retain the epoch with the highest F1 score on the validation set. As robustness checks, we also fine-tune (i) a version of the model with half of its layers frozen and (ii) the well-known BERTweet model \citep{nguyen2020bertweet}. Both alternatives perform slightly worse, but their outputs remain highly correlated with those of our baseline classifier, with Matthews correlation coefficients of 0.93 and 0.85, respectively.}

The fine-tuned classifier assigns each tweet a probability distribution over the three classes---blame, credit, and none---and labels the tweet according to the highest probability class. Ambiguous cases, in which the model assigns similar probabilities to multiple classes, are extremely rare.\footnote{\Cref{fig:p_dists} shows that, conditional on a predicted label, the associated probability distributions are strongly left-skewed and concentrated near 1, indicating that the classifier is rarely uncertain.} On the validation set, the model achieves an accuracy of 0.83, an F1 score of 0.84, and a Matthews correlation coefficient of 0.73, indicating strong performance by the standards of comparable text-classification tasks.

A natural concern is that our construction might miss a broader category of causal rhetoric that links political actions to outcomes without taking either a negative or a positive stance, and therefore extends beyond blame and credit. In practice, however, such neutral-toned causal tweets are exceedingly rare in our hand-labeled data, accounting for just 0.4 percent of observations. This suggests that, as a qualitative reading of political discourse would also imply, politicians almost never use causal rhetoric in a neutral way. Instead, when they make such causal claims, they typically do so to shape the attribution process through claims of blame or credit. To further validate this point, we train a separate classifier using only the binary causal label from the annotated dataset and find that its output is highly correlated with an indicator for whether a tweet is classified as either blame or credit ($\rho = 0.9$). Taken together, these patterns confirm that causal rhetoric in politics consists almost entirely of assigning blame and claiming credit.

\subsection{Validation}\label{sec:validation}

We validate the measure against both external benchmarks and internal linguistic patterns. The external validation compares our labels with independent classifications from America's Political Pulse. The internal validation asks whether tweets classified as blame and credit exhibit the semantic, syntactic, and emotional patterns one would expect from these categories.

\paragraph*{External validation} 
We begin by validating our measure against America's Political Pulse, a project led by the Polarization Research Lab that independently classifies congressional communications into rhetorical categories \citep{Westwood2024}. The data cover speeches, newsletters, tweets, press releases, and public statements beginning in 2023 and include categories such as policy-attack and credit-claiming. For comparability with our analysis, we restrict the data to tweets posted through the end of our sample period, yielding 158,745 tweets from 449 legislators.

These categories are not identical to ours, but they provide a natural benchmark. Policy-attack is close to blame because both capture negative rhetoric directed at the actions of political actors. Credit-claiming is close to credit because both capture positive rhetoric that associates favorable outcomes or accomplishments with political actors.\footnote{In their codebook, \citet{Westwood2024} define policy-attack as ``communications about objecting to or raising concerns about a specific policy, law, or court ruling; using fact-based arguments even if critical or negative; avoiding emotional appeals, inflammatory language, claims of extremism, or personal attacks on individuals involved with the policy, including accusations of lying or withholding information,'' and credit-claiming as ``communications about creating or passing legislation; securing government spending, grants, or funding; or emphasizing personal or party accomplishments in office.''}

The two measures align closely at the tweet level. For each tweet we can match across datasets, we compare our label with the corresponding America's Political Pulse classification. Over the common sample period, this procedure yields 101,689 matched tweets. For blame, our label agrees with the policy-attack label in 87 percent of matched tweets, with a correlation of 0.62 between the two binary indicators. For credit, our label agrees with the credit-claiming label in 84 percent of matched tweets, with a corresponding correlation of 0.60.

The measures also align in the cross-section of politicians. This validation is important because several of our empirical exercises use politician-level variation in the intensity of blame and credit. We restrict attention to the 443 politicians observed in both datasets over the common sample period, from the beginning of 2023 through the end of our sample. For each politician, we compute the share of tweets classified as policy-attack and credit-claiming in America's Political Pulse and correlate these shares with our corresponding shares of blame and credit tweets. \Cref{fig:validation} shows strong positive correlations: 0.85 between blame and policy-attack, and 0.78 between credit and credit-claiming. The tweet-level and politician-level comparisons therefore provide strong external support for the validity of our measure.

\begin{figure}[t]
    \centering
    \caption{External Validation against America's Political Pulse}
    \label{fig:validation}
    \begin{subfigure}{0.4\textwidth}
        \caption{Blame}
        \label{fig:validation_blame}
        \includegraphics[width=\linewidth]{inputs_paper/current/figures/validation_blame.pdf}
    \end{subfigure}
    \hspace{1cm}
    \begin{subfigure}{0.4\textwidth}
        \caption{Credit}
        \label{fig:validation_merit}
        \includegraphics[width=\linewidth]{inputs_paper/current/figures/validation_merit.pdf}
    \end{subfigure}
    \vspace{0.3cm}
    \caption*{\scriptsize \textit{Notes}: This figure presents binned scatterplots comparing our politician-level measures of blame and credit with the corresponding politician-level measures from America's Political Pulse \citep{Westwood2024}. In Panel (a), the horizontal axis reports the share of tweets classified as blame in our data, while the vertical axis reports the share of tweets classified as policy-attack in America's Political Pulse. Panel (b) is analogous for credit and credit-claiming. Both sources are restricted to the common sample period, from the beginning of 2023 through the end of our sample. All variables are standardized. Observations correspond to politicians and are grouped into 100 bins. The correlation coefficient is reported in the upper-left corner of each panel.}
\end{figure}

\paragraph*{Internal validation} We next turn to internal validation exercises, which help clarify the substantive content captured by our labels. We begin by examining the targets of causal claims. Syntactically, we measure whether tweets are self- or other-referential by comparing the relative frequency of second- and third-person pronouns to first-person pronouns. \Cref{fig:linguistic_features_selfother} shows that blame tweets are more outward-directed, framing responsibility around others, whereas credit tweets are more self-directed. Semantically, we classify whether the attribution targets Democrats or Republicans.\footnote{Target classification is based on the Political DEBATE language model developed by \citet{burnham2025political}. Political DEBATE is specialized in zero-shot and few-shot classification of political documents, with performance on par with, or better than, state-of-the-art language models.} Restricting attention to the 40 percent of tweets for which the target can be identified as either Democratic or Republican, \Cref{fig:linguistic_features_entities} shows that blame is disproportionately directed at the opposing party, whereas credit is more often directed toward one's own side.

We then examine the temporal orientation of causal rhetoric. Although our concept of causal rhetoric is not inherently temporal, blame and credit may differ in whether they connect actions to realized outcomes or to expected consequences. We capture this distinction using normalized counts of past and future tense. \Cref{fig:linguistic_features_tense} shows that blame is predominantly retrospective, whereas credit is more forward-looking, consistent with findings from psychology \citep{malle2014theory}.

We next study the emotional content associated with each rhetorical style. Using a RoBERTa model fine-tuned for emotion detection in tweets \citep{camacho2022tweetnlp}, we decompose the emotional profile of blame and credit tweets. \Cref{fig:emotions} shows that blame is dominated by anger, followed by disgust and fear, consistent with the idea that anger is directional and closely tied to causal attribution \citep{lazarus1991cognition}. Credit, by contrast, is mainly associated with optimism, reinforcing the earlier result that credit is more forward-looking.

Finally, we examine diagnostic language. Following \citet{gentzkow2010drives}, we extract the bigrams that are most distinctive of each category. As shown in \Cref{fig:distinctiveness_bigrams_dems,fig:distinctiveness_bigrams_reps}, these bigrams closely match both intuition and the linguistic patterns documented above.\footnote{For Democrats, ``trump administration'' is diagnostic of blame, while ``will help'' is diagnostic of credit. For Republicans, ``southern border'' characterizes blame, while ``will help'' again identifies credit.} We complement these diagnostics with illustrative tweets in \Cref{tab:sample_tweets_dems,tab:sample_tweets_reps}.

As an additional check, our measure is not reducible to existing text-based measures of political language. \Cref{fig:coefs_text} compares blame and credit with sentiment, emotionality, and moral terminology, three dimensions commonly used to characterize political communication. The correlations are intuitive but limited: blame is more negative and credit more positive, but neither category is simply a relabeling of sentiment, emotional language, or moral language. This evidence supports the interpretation of blame and credit as a distinct dimension of political rhetoric.

\medskip
This procedure yields the measure of causal rhetoric used throughout the rest of the paper. Starting from a definition of causal rhetoric as language linking political actions to outcomes, we fine-tune a language model to classify each tweet as blame, credit, or neither. The classifier performs well against hand-coded labels, aligns with an independent external benchmark, and displays the linguistic patterns expected of causal language in politics. We therefore use this measure as the basis for the empirical analysis that follows.

\section{The Supply of Causal Rhetoric}\label{sec:supply}

We use our measure of blame and credit to document three facts about the supply of causal rhetoric in congressional tweets. First, causal rhetoric is a large part of politicians' communication, not a rare rhetorical device. Second, it is the language of political disagreement: when politicians assign blame or claim credit, the ruling party and the opposition diverge much more sharply in how they evaluate the political environment than they do in other tweets. Third, causal rhetoric has become substantially more common over time. These facts motivate the rest of the paper by showing that blame and credit are quantitatively important, tied to political disagreement, and increasingly prominent in elite political communication.

\subsection{Prevalence}

Causal rhetoric accounts for a substantial share of congressional communication on social media. In the full corpus, 16 percent of tweets are classified as blame and 20 percent as credit. Taken together, more than one third of congressional tweets assign blame or claim credit. This makes causal rhetoric a major margin of politicians' communication rather than a narrow class of exceptional messages.

This prevalence is not evenly distributed across issues. \Cref{fig:topic_blamemerit} shows that blame and credit are especially common in policy domains rather than sociocultural ones.\footnote{We classify tweets into eight broad topics using the Political DEBATE language model by \cite{burnham2025political}: economy, environment, healthcare, and immigration on the policy side; and gender, gun control, policing, and racial relations on the sociocultural side.} Causal rhetoric is most prevalent in topics such as the economy, healthcare, immigration, and the environment, where political actors can most naturally be linked to social and economic outcomes, and less prevalent in topics such as gender, gun control, policing, and racial relations. This pattern is consistent with the interpretation of blame and credit as forms of causal attribution: politicians use them most where political actions can be connected to observable consequences.

Causal rhetoric is not concentrated among a narrow demographic type of politician. \Cref{fig:coefs_author} relates rhetorical style to politician characteristics. Restricting attention to demographic characteristics, age, gender, race, and education have little explanatory power for either blame or credit. Causal rhetoric therefore appears to be a broad feature of congressional communication rather than a style used only by a specific demographic group of politicians.

\subsection{Disagreement}

Causal rhetoric is also the language of political disagreement. We focus on a specific form of disagreement: divergence in how politicians evaluate the political environment. To measure this disagreement, we compare the average standardized sentiment of tweets posted by members of the ruling party with that of tweets posted by members of the opposition. This sentiment gap captures whether politicians on opposite sides of political control describe the state of the world in systematically different evaluative terms. We compute the gap separately for all tweets, tweets classified as blame or credit, and tweets classified as neither blame nor credit.

\Cref{fig:sentiment_divide_pooled} shows that the sentiment gap is concentrated in causal tweets. Across all tweets, the ruling-party-versus-opposition gap is 0.33 standard deviations. Among tweets classified as blame or credit, the gap nearly doubles, reaching 0.64 standard deviations. Among other tweets, it is only 0.13 standard deviations. Political disagreement is therefore not expressed equally across all communication. It is especially concentrated in causal rhetoric, where politicians connect political actions to outcomes through blame and credit.

\begin{figure}[t]
    \centering
    \caption{Causal Rhetoric and Disagreement}
    \label{fig:sentiment_divide_pooled}
    \includegraphics[width=0.55\linewidth]{inputs_paper/current/figures/sentiment_divide_pooled.pdf}
    \caption*{\scriptsize \textit{Notes}: The figure reports evaluative disagreement, measured as the difference in average standardized sentiment between tweets posted by members of the ruling party and tweets posted by members of the opposition. Higher values indicate a larger divergence in how politicians on opposite sides of political control evaluate the political environment. We compute this measure separately for all tweets, tweets classified as blame or credit, and tweets classified as neither blame nor credit. Bars represent 95 percent confidence intervals.\par}
\end{figure}

This concentration of political disagreement in causal tweets is not mechanical. Blame and credit tweets, by construction, refer to political actors more often than non-causal tweets, so one concern is that they may exhibit larger sentiment gaps simply because they more often identify political targets. \Cref{fig:sentiment_divide_target_pooled} addresses this concern by restricting the analysis to tweets that can be clearly identified as targeting either Democrats or Republicans. The same pattern holds in this restricted sample, although the sentiment gaps are higher in levels. Causal rhetoric therefore appears to be the language of political disagreement, not merely a by-product of politicians mentioning political actors.

\subsection{Trends}

The prevalence of causal rhetoric, already large at the beginning of the sample, has grown over time. \Cref{fig:time_trend} plots the yearly share of tweets classified as blame and credit from 2012 to 2023. Both dimensions rise substantially, from about 10 percent of tweets at the beginning of the sample to more than 20 percent by the end. Taken together, blame and credit grow from roughly one fifth to nearly one half of all congressional tweets.

The timing of the increase differs somewhat across the two rhetorical styles. Blame rises most sharply between 2016 and 2017, increasing by more than 10 percentage points in a single year. Credit follows a similar trajectory, although its growth is more gradual and concentrated between 2017 and 2020. By 2020, both series appear to plateau at high levels. Thus, an already large component of congressional communication becomes substantially more prominent over the sample period.

The rise does not appear to be driven by mechanical changes in the composition of the data. A first possibility is that different politicians are speaking, or that politicians are speaking about different issues. \Cref{fig:decomposition} addresses this concern by estimating yearly changes in blame and credit while progressively adding politician and topic fixed effects, using the topic classification described above. The upward trend remains after accounting for both sets of fixed effects, indicating that the increase occurs within politicians and within topics rather than merely reflecting congressional turnover or shifts in the issues politicians discuss.

\begin{figure}[t]
    \centering
    \caption{Causal Rhetoric over Time}
    \label{fig:time_trend}
    \includegraphics[width=0.65\linewidth]{inputs_paper/current/figures/time_both.pdf}
    \caption*{\scriptsize \textit{Notes}: The figure presents the yearly share of tweets classified as blame and credit.}
\end{figure}

Additional checks address other mechanical explanations. The pattern is not driven by changes in sample coverage: \Cref{fig:time_trend_nosenators} shows similar trends when we exclude Senators, whose tweets enter our data only in 2017. It is also not a generic change in political language on Twitter: \Cref{fig:time_trend_comparison} shows that sentiment, emotionality, and moral language do not exhibit comparable increases. Nor is the rise explained by Twitter platform changes. Appendix \ref{app:platform} shows no discontinuous increase around either the introduction of algorithmic feed curation in 2016 or the expansion of the character limit in 2017. Finally, Appendix \ref{app:trump} shows that the rise is not simply a Trump-style shift in congressional communication: Trump's distinctive rhetorical feature is non-causal negative sentiment, not unusually high blame or credit.

\medskip
Taken together, the supply evidence shows that causal rhetoric is a major and increasingly prominent component of politicians' communication on social media. It is common, is the language of political disagreement, and increases over time. The next section therefore asks whether this rhetoric generates political returns.

\section{The Political Returns of Causal Rhetoric}\label{sec:returns}

We next ask whether causal rhetoric generates political returns. We interpret politicians' use of blame and credit on social media as a rhetorical production problem \citep{aridor2024economics}: politicians choose rhetorical inputs partly based on the political returns those inputs can generate. This section studies those returns by asking whether exposure to politicians assigning blame and claiming credit increases political support.

We study this question using two complementary sources of evidence. The first is small campaign contributions. Although not perfectly representative of the general population, small donations can still provide a revealed-preference measure of political support that can be linked systematically to politicians' communication in our data \citep{petrova2021social,bouton2022small}. They also allow us to distinguish two margins of support: whether rhetoric mobilizes additional donors, and whether it deepens support by increasing average giving among donors.

Using a cross-county shift-share-style exposure design that combines politician-month rhetoric with quasi-exogenous geographic variation in Twitter adoption \citep{muller2023hashtag}, we estimate the effect of exposure to politicians' blame- and credit-intensive communication on small-donor support. The design asks whether counties more exposed to Twitter respond more to the same politician in months when that politician uses more blame or credit, rather than considering a timing comparison around individual tweets. The estimates show asymmetric returns. Blame has the larger aggregate return, mainly by expanding the donor base. Credit generates much smaller aggregate returns, with detectable effects only on average giving. These effects also differ across donor types: ideologically extreme donors respond more to blame, while ideologically moderate donors respond more to credit.

The second source of evidence is survey data on voting intentions in the 2020 House elections. The survey analysis complements the donation design in three ways. It measures support directly as intended vote choice, covers respondents beyond the selected population of donors, and uses a short-run timing comparison.

Using an event-study design around blame-only and credit-only days before the election, we examine whether support for a politician changes after that politician's Twitter communication consists entirely of one rhetorical style. Support increases in the week after blame-only days, especially among respondents who report using social media for political news. Credit-only days show no comparable effect. The survey evidence therefore points in the same direction as the donation evidence while also supporting the social-media exposure channel.

\subsection{Empirical Strategy for Donations}\label{sec:donations-identification}

We estimate the effect of causal rhetoric on small-donor support using a shift-share exposure design that relies on quasi-exogenous exposure shares \citep{goldsmith2020bartik, borusyak2025practical}. The shifts are politician-month rhetoric: in a given month, a politician uses more or less blame or credit. The exposure share is county-level exposure to Twitter: residents of counties with higher Twitter adoption are more likely to encounter the same politician's Twitter communication. Identification comes from within-politician-month comparisons: we ask whether counties with greater quasi-exogenous Twitter exposure donate more to the same politician in months when that politician uses more blame or credit, net of persistent politician-county ties and county-month shocks.

We use the geographic detail in campaign finance records to construct a politician-by-county-by-month panel. For each politician, county, and month in which the politician tweets, we measure three outcomes from small donors: total donation revenue, the number of donors, and the average amount contributed per donor. We estimate:
\begin{align}
    y_{imc} =& \ \beta_1 (\text{Blame}_{im} \times \text{Users}_{c}) + \beta_2 (\text{Credit}_{im} \times \text{Users}_{c}) \label{reg:causal_spec} \\
    & + \gamma X_{imc} + \lambda_{ic} + \mu_{im} + \eta_{mc} + \varepsilon_{imc}. \notag
\end{align}
Here, $y_{imc}$ is the outcome for politician $i$ in month $m$ and county $c$: log(1+donation revenue), log(1+number of donors), or log(1+average per donor). $\text{Blame}_{im}$ and $\text{Credit}_{im}$ denote the standardized shares of blame and credit tweets posted by politician $i$ in month $m$. $\text{Users}_{c}$ measures county-level Twitter use, defined as the log of one plus the number of Twitter users in county $c$. The vector $X_{imc}$ denotes controls. In our preferred specification, it includes $\text{Volume}_{im} \times \text{Users}_{c}$, where $\text{Volume}_{im}$ is the standardized total number of tweets posted by politician $i$ in month $m$. This separates exposure to rhetorical style from exposure to overall political activity on Twitter.

The fixed effects make the exposure comparison demanding. Politician-by-month fixed effects $(\mu_{im})$ absorb the politician-month rhetorical shift itself, as well as any month-specific shock to politician $i$, including changes in popularity, fundraising effort, or national visibility that affect donations across all counties. Politician-by-county fixed effects $(\lambda_{ic})$ absorb persistent differences in politician $i$'s support across counties, including geographic ties between politicians and specific places. County-by-month fixed effects $(\eta_{mc})$ absorb local shocks to donations in month $m$ that are common across politicians, such as county-level changes in political engagement or economic conditions. The coefficients are therefore identified by comparing, within the same politician-month, whether donations respond more in counties with higher Twitter exposure, net of persistent politician-county links and county-month shocks.

The main identification challenge is that county Twitter adoption may be endogenous to local political and economic conditions. Counties with greater partisan competition, higher donor capacity, or different unobservable characteristics may exhibit different levels of Twitter adoption and may also differ systematically in political engagement. If so, the interaction between politicians' rhetoric and county Twitter use could partly capture differences in the local demand for political communication rather than the causal effect of exposure to that rhetoric. To address this concern, we exploit quasi-exogenous variation in Twitter adoption induced by the platform's early diffusion after the South by Southwest (SXSW) festival in 2007, a well-documented shock that accelerated Twitter's initial spread in the United States \citep{muller2023hashtag, fujiwara2024effect}. Specifically, we instrument county-level Twitter adoption with the number of followers of the official SXSW Twitter account who joined Twitter in 2007. The instrument isolates variation in local Twitter use generated by early exposure to the platform, rather than by contemporaneous local determinants of political donations.\footnote{This instrument is now widely used in the literature \citep[e.g.,][]{boken2026returns}. We refer to the original papers for detailed evidence on both its relevance for the diffusion of local Twitter activity and its plausible exogeneity.}

In practice, we instrument $\text{Users}_{c}$ with log$(1+\text{SXSWFollowers2007}_{c})$, where SXSW\-Followers2007\textsubscript{$c$} denotes the number of followers of the official SXSW account in county $c$ who joined Twitter in 2007, after the SXSW Festival. We estimate a standard two-stage least squares specification, instrumenting each interaction in Equation \ref{reg:causal_spec} that involves county Twitter adoption with the corresponding interaction based on SXSW\-Followers2007\textsubscript{$c$}. Following \citet{muller2023hashtag, fujiwara2024effect}, we also interact our main regressors of interest with the number of SXSW followers who joined Twitter before 2007, at any point in 2006. This ensures that identification comes from comparing counties that are similar in their pre-existing propensity to engage with SXSW, and therefore more comparable in their baseline relationship to the platform.

The identifying assumption is a share-exogeneity condition. Conditional on the fixed effects, tweet-volume exposure, and pre-2007 SXSW propensity controls, counties with greater SXSW-induced Twitter adoption would have experienced similar changes in politician-specific donations across more and less blame-intensive, or more and less credit-intensive, months absent differential exposure to politicians' Twitter communication. The design does not require politicians' rhetorical choices to be randomly timed: politicians may choose when to assign blame or claim credit in response to political events. Identification comes instead from the SXSW-induced component of county Twitter exposure, which must be unrelated to unobserved county-level donation responses to politician-month rhetorical shifts except through exposure to the politician's tweets. Under this assumption, the estimates identify the causal effect of exposure to more blame- or credit-intensive politician-month communication periods on small-donor support.

\subsection{Results for Donations}\label{sec:donations-results}

Exposure to causal rhetoric increases small-donor support, but the returns are asymmetric. \Cref{tab:donations-main} reports estimates of Equation \ref{reg:causal_spec} using log total small-donation revenue as the outcome in column (1).\footnote{\Cref{tab:donations-fs}, \Cref{tab:donations-ols}, and \Cref{tab:donations-rf} report the corresponding first-stage, OLS, and reduced-form results.} A one-standard-deviation increase in the monthly share of blame tweets increases donation revenue by roughly 2.6 percent in a county with average Twitter adoption. The corresponding increase for credit is about 0.5 percent. Blame therefore generates the larger aggregate return.

\begin{table}[t]
\caption{Donations}
\label{tab:donations-main}
\centering
\begin{footnotesize}
\begin{adjustbox}{max width=\textwidth}
\begin{tabular}{l*{3}{c}}
\toprule
& (1) & (2) & (3) \\
\midrule
& \makecell{Revenue\\from Donations} & \makecell{Number\\of Donors} & \makecell{Average\\per Donor} \\
\midrule
Blame x Users&0.005\sym{***}&0.003\sym{***}&0.003\sym{***}\\
&(0.001)&(0.000)&(0.001)\\
Credit x Users&0.001\sym{***}&0.000\sym{**}&0.001\sym{***}\\
&(0.000)&(0.000)&(0.000)\\
Blame x SXSWFollower2006&-0.006&-0.002&-0.004\\
&(0.004)&(0.002)&(0.003)\\
Credit x SXSWFollower2006&0.003&0.001&0.002\sym{*}\\
&(0.002)&(0.001)&(0.001)\\
\midrule
Politician x Month FE & \checkmark & \checkmark & \checkmark  \\
Politician x County FE & \checkmark & \checkmark & \checkmark  \\
County x Month FE & \checkmark & \checkmark & \checkmark  \\
Tweet Volume Control & \checkmark & \checkmark & \checkmark  \\
\midrule
Observations &   168,232,932 &   168,232,932 &   168,232,932 \\
Clusters &         3,108 &         3,108 &         3,108 \\
F statistic & 63.65 & 63.65 & 63.65 \\
\bottomrule
\end{tabular}\end{adjustbox}
\\[1em]
\caption*{\scriptsize \textit{Notes}: The table presents 2SLS estimates of Equation \ref{reg:causal_spec}. The outcomes are log-donation revenue in column (1), log-number of donors in column (2), and log-average donation per donor in column (3). The main regressors are the standardized monthly shares of blame and credit tweets posted by the politician, each interacted with the log of one plus the number of Twitter users in the county. The table also reports the coefficients on the corresponding interactions with the log of one plus the number of SXSW followers in the county who joined Twitter in 2006. Interactions with county Twitter users are instrumented using the corresponding interactions with the log of one plus the number of SXSW followers in the county who joined Twitter in 2007. All specifications include politician-by-month, politician-by-county, and county-by-month fixed effects, as well as the interaction between tweet volume and county Twitter users. Standard errors in parentheses are clustered at the county level. *, **, *** denote significance at the 10\%, 5\%, and 1\% levels, respectively.\par}
\end{footnotesize}
\end{table}

These aggregate returns operate through different margins of support. Donation revenue can increase because rhetoric mobilizes additional donors, raising the number of donors, or because it deepens support among donors, raising the average amount contributed per donor. To distinguish between these margins, \Cref{tab:donations-main} uses the log number of donors and log average donation per donor as outcomes in columns (2) and (3). Blame is the only rhetorical style that expands the donor base. A one-standard-deviation increase in blame increases the number of donors by about 1.6 percent in a county with average Twitter adoption, while credit has essentially no effect on this margin. Credit instead has a smaller effect on support deepening: a one-standard-deviation increase in credit raises the average donation per donor by about 0.5 percent in a county with average Twitter adoption. The corresponding estimates for blame on average donation size are positive but less precise across some robustness checks. Overall, blame is effective at mobilizing additional donors, while credit only slightly increases average giving among those already willing to donate.

\paragraph*{Persuasion rate} 

To benchmark the magnitude of the donor-mobilization effect, we compute persuasion rates following \citet{dellavigna2010persuasion}, focusing on the donor-count outcome.\footnote{As emphasized by \citet{boken2026returns}, persuasion rates are most naturally interpreted when the underlying outcome is close to a binary participation decision.} The implied persuasion rate is 0.94\%.\footnote{Appendix \ref{app:persuasion} details the construction of this estimate and shows that it should be interpreted as a lower bound.} This magnitude is comparable to estimates for political advertising \citep{spenkuch2018political} and opening a Twitter account \citep{petrova2021social}, though smaller than the effects associated with virality \citep{boken2026returns} and the average persuasion rates summarized by \citet{dellavigna2010persuasion}.

\paragraph*{Heterogeneity by donor ideology}

The donation returns to blame and credit differ not only by margin of support, but also by donor ideology. The aggregate results show that blame is more effective at mobilizing additional donors, while credit slightly increases average giving. We next ask whether these two rhetorical styles activate different ideological constituencies within the donor base.

To study this heterogeneity, we use the donor-level ideology measure constructed by \citet{bonica2014mapping} from campaign finance records, which is analogous to NOMINATE scores for legislators. Within each electoral cycle, we classify donors as moderate or extreme depending on whether the absolute value of their ideological score falls below or above the cycle-specific median. We then re-estimate Equation \ref{reg:causal_spec} separately by donor type.

\Cref{tab:donations-het} reveals a clear ideological divide. Moderate donors respond positively to credit and negatively to blame. Extreme donors display the opposite pattern: they respond positively to blame and negatively to credit. This symmetry is qualitative rather than quantitative. The positive response of extreme donors to blame is substantially larger than the positive response of moderate donors to credit. These differences indicate that blame and credit do not merely operate through different margins of support; they also activate different ideological constituencies within the donor base. The two rhetorical styles are therefore politically complementary despite the larger aggregate returns to blame.

\begin{table}[t]
\caption{Donations by Donor Ideology}
\label{tab:donations-het}
\centering
\begin{footnotesize}
\begin{adjustbox}{max width=\textwidth}
\begin{tabular}{l*{6}{c}}
\toprule
& (1) & (2) & (3) & (4) & (5) & (6) \\
\midrule
 & \multicolumn{3}{c}{Moderate Donors} & \multicolumn{3}{c}{Extreme Donors}\\
\cmidrule(lr){2-4} \cmidrule(lr){5-7}
& \makecell{Revenue\\from Donations} & \makecell{Number\\of Donors} & \makecell{Average\\per Donor} & \makecell{Revenue\\from Donations} & \makecell{Number\\of Donors} & \makecell{Average\\per Donor} \\
\midrule
Blame x Users&-0.003\sym{***}&-0.002\sym{***}&-0.001\sym{***}&0.010\sym{***}&0.005\sym{***}&0.006\sym{***}\\
&(0.001)&(0.000)&(0.000)&(0.002)&(0.001)&(0.001)\\
Credit x Users&0.005\sym{***}&0.002\sym{***}&0.003\sym{***}&-0.003\sym{***}&-0.002\sym{***}&-0.002\sym{***}\\
&(0.001)&(0.000)&(0.000)&(0.001)&(0.000)&(0.000)\\
Blame x SXSWFollower2006&-0.001&-0.000&-0.000&-0.005&-0.002&-0.003\\
&(0.003)&(0.001)&(0.002)&(0.006)&(0.003)&(0.004)\\
Credit x SXSWFollower2006&0.000&-0.000&0.000&0.003&0.001&0.002\sym{*}\\
&(0.003)&(0.001)&(0.002)&(0.002)&(0.001)&(0.001)\\
\midrule
Politician x Month FE & \checkmark & \checkmark & \checkmark & \checkmark & \checkmark & \checkmark  \\
Politician x County FE & \checkmark & \checkmark & \checkmark & \checkmark & \checkmark & \checkmark  \\
County x Month FE & \checkmark & \checkmark & \checkmark & \checkmark & \checkmark & \checkmark  \\
Tweet Volume Control & \checkmark & \checkmark & \checkmark & \checkmark & \checkmark & \checkmark  \\
\midrule
Observations &   168,232,932 &   168,232,932 &   168,232,932 &   168,232,932 &   168,232,932 &   168,232,932 \\
Clusters &         3,108 &         3,108 &         3,108 &         3,108 &         3,108 &         3,108 \\
F statistic & 63.65 & 63.65 & 63.65 & 63.65 & 63.65 & 63.65 \\
\bottomrule
\end{tabular}\end{adjustbox}
\\[1em]
\caption*{\scriptsize \textit{Notes}: The table presents 2SLS estimates of Equation \ref{reg:causal_spec} separately by donor ideology. Columns (1)--(3) use outcomes computed from ideologically moderate donors, and columns (4)--(6) use outcomes computed from ideologically extreme donors. Within each donor group, the outcomes are log-donation revenue, log-number of donors, and log-average donation per donor. The main regressors are the standardized monthly shares of blame and credit tweets posted by the politician, each interacted with the log of one plus the number of Twitter users in the county. The table also reports the coefficients on the corresponding interactions with the log of one plus the number of SXSW followers in the county who joined Twitter in 2006. Interactions with county Twitter users are instrumented using the corresponding interactions with the log of one plus the number of SXSW followers in the county who joined Twitter in 2007. All specifications include politician-by-month, politician-by-county, and county-by-month fixed effects, as well as the interaction between tweet volume and county Twitter users. Standard errors in parentheses are clustered at the county level. *, **, *** denote significance at the 10\%, 5\%, and 1\% levels, respectively.\par}
\end{footnotesize}
\end{table}

\paragraph*{Robustness checks} 

We estimate several additional specifications that address alternative interpretations of the donation results. First, the estimated returns to blame and credit may reflect exposure to negative and positive sentiment rather than exposure to causal attribution. This would be a concern if politicians systematically use blame when they also use more non-causal negative rhetoric, or credit when they also use more non-causal positive rhetoric. In that case, the estimates could partly capture the returns to affective sentiment rather than the returns to causal rhetoric. To address this concern, we control for the shares of non-causal tweets with negative sentiment and non-causal tweets with positive sentiment, each interacted with county Twitter adoption and instrumented with the corresponding SXSW interaction. The results, reported in \Cref{tab:donations-robustness}, are essentially unchanged.

Second, the estimates may reflect exposure to partisan communication more generally. This would be a concern if politicians use blame and credit disproportionately during moments of partisan conflict, when they are also more likely to criticize the opposing party or praise their own side. In that case, the estimates could partly capture broader shifts in partisan rhetoric rather than the returns to assigning blame or claiming credit. To address this concern, we construct measures of non-causal attacks and non-causal praise. Non-causal attacks are tweets that are non-causal, negative in tone, and about the opposing party; non-causal praises are tweets that are non-causal, positive in tone, and about the politician's own party. We interpret these measures as coarse proxies for correlated partisan communication. We then control for the shares of these tweets, interacted with county Twitter adoption and instrumented with the corresponding SXSW interactions.\footnote{To identify whether tweets are about the opposing party or the politician's own party, we use the same target-classification procedure employed in the validation exercises, based on the Political DEBATE language model by \citet{burnham2025political}.} The results, also reported in \Cref{tab:donations-robustness}, are essentially unchanged.

Third, blame and credit may coincide with other politically salient content. Politicians may assign more blame or claim more credit when economic news is released, scandals emerge, or other salient information becomes public. If county responsiveness to this accompanying content is correlated with SXSW-induced Twitter adoption through demographics, media-consumption patterns, or socioeconomic conditions, the estimates could partly capture differential responses to correlated political events rather than the rhetorical margin itself. To address this concern, we estimate a richer specification that allows the effects of blame, credit, and tweet volume to vary flexibly with county characteristics. We use the 29 county-level covariates employed by \citet{fujiwara2024effect}, summarizing them into eight principal components to avoid a high-dimensional set of interactions, and interact these principal components with blame, credit, and tweet volume.\footnote{The 29 covariates include demographic and socioeconomic characteristics, as well as measures of media exposure such as prime-time television viewing and Fox News viewership. \Cref{tab:controls} lists the underlying covariates. We retain principal components using the Kaiser criterion, keeping components with eigenvalues greater than one. This yields eight components, which together explain 74\% of the variance in the 29 county-level covariates.} This specification absorbs differential responsiveness associated with observable county characteristics, so correlated political events would have to be differentially amplified through the SXSW-induced Twitter-exposure margin over and above these demographic, socioeconomic, media-consumption, and political-engagement channels. The results, also reported in \Cref{tab:donations-robustness}, are essentially unchanged, except for blame becoming noisier on average donation.

Finally, the log transformations may be sensitive to the large number of zeros in the donation outcomes \citep{chen2024logs}. We therefore re-estimate the same specification, as well as the previous robustness checks, using a binary outcome equal to one if a politician receives at least one donation from a given county in a given month, and zero otherwise. The results, reported in \Cref{tab:donations-any}, are consistent with the donor-count evidence: the effects point in the same direction even when we use a coarser and more restrictive measure of donor mobilization that captures only whether the politician receives any donation from a county-month, rather than how many donors contribute.

\medskip
Overall, the donation evidence points to asymmetric political returns. Blame increases small-donor support primarily by mobilizing additional donors. Credit generates smaller returns and only increases average giving among donors. These results establish the donation margin of political support. We next ask whether the same pattern appears in a broader survey measure of intended vote choice.

\subsection{Empirical Strategy for Voting Intentions}\label{sec:survey-identification}

We estimate whether causal rhetoric changes voting intentions using an event-study design around days when a politician's Twitter communication consists entirely of a single rhetorical style. For each politician, we define a blame-only day as a day on which all the politician's tweets are classified as blame; credit-only days are defined analogously. This definition isolates days with a clear rhetorical style: the politician uses only blame or only credit. The identifying comparison asks whether support for the politician changes after blame-only or credit-only days, net of district-month conditions, date shocks, partisan alignment, and respondent characteristics.

We use Nationscape survey responses on voting intentions in the 2020 House elections. Our main outcome, $\text{Support}_{idt}$, is an indicator equal to one if respondent $i$ in district $d$, interviewed on date $t$, reports intending to vote for the politician, and zero otherwise. We link each respondent to the Twitter communication of the elected representative serving the respondent's congressional district. We then compare respondents interviewed in the seven days after blame-only or credit-only days with respondents interviewed during the six days before the rhetoric-only day and on the rhetoric-only day itself. Respondents interviewed on the rhetoric-only day itself are included in the comparison group because we do not observe whether the survey response occurred before or after the politician posted on Twitter.

Formally, for each rhetorical style $r \in \{\text{Blame}, \text{Credit}\}$, we construct an indicator $\text{Post-$r$}_{dt}$ equal to one if respondent $i$ is interviewed in district $d$ during the seven days after a rhetoric-only day of style $r$ by the district's politician. The comparison group consists of respondents interviewed during the six days before the rhetoric-only day and on the rhetoric-only day itself. We exclude observations that fall in both the pre- and post-period for the same rhetorical style. We then estimate:
\begin{align}
    \text{Support}_{idt} = \beta \ \text{Post-$r$}_{dt}
    + \lambda_{d \times m(t)}
    + \mu_t
    + \eta_{a(i,d)}
    + \theta_{X_i}
    + \varepsilon_{idt},
    \label{reg:survey}
\end{align}
where $\text{Support}_{idt}$ is the voting-intention outcome; $\lambda_{d \times m(t)}$ are district-by-month fixed effects; $\mu_t$ are date fixed effects; $\eta_{a(i,d)}$ are fixed effects for respondent--politician partisan alignment, defined by respondent party identification interacted with an indicator for whether the politician is Republican; and $\theta_{X_i}$ denotes fixed effects for respondent gender, age group, race, education, and employment status.\footnote{Respondent party identification is coded in four categories: Democrat, Republican, Independent, and something else.} Standard errors are clustered at the congressional-district level.

The identifying assumption is that, within a district and month, the exact timing of survey interviews is as-good-as-random with respect to the politician's blame-only or credit-only days, after accounting for date effects, partisan alignment, and respondent demographics. District-by-month fixed effects compare respondents in the same district and month, holding fixed local political conditions and the local campaign environment. Date fixed effects absorb national shocks common to all districts. Party-identification-by-politician-party fixed effects capture partisan alignment by allowing baseline support to differ across respondent party groups depending on whether the politician is Democratic or Republican. Respondent-characteristic fixed effects absorb differences in support across observable demographic groups.

The balance checks support this timing comparison. \Cref{tab:survey-balance-demographics} shows that observable demographic characteristics of respondents interviewed before rhetoric-only days are statistically indistinguishable from those of respondents interviewed after rhetoric-only days. \Cref{tab:survey-balance-ideology} shows the same for partisan alignment, measured by the interaction between respondent party identification and the politician's party. These checks reduce the concern that the before-after comparison is driven by changes in the composition of respondents within district-months.

A further concern is that rhetoric-only days may be isolated episodes or concentrated among a small subset of politicians. We show that neither is the case. \Cref{fig:survey-rhetoric-only-frequency} shows that rhetoric-only days are recurrent: they account for 10.6\% of all politician-days with at least one tweet and 15.3\% of days with at least one blame or credit tweet, with both blame and credit contributing. \Cref{fig:survey-rhetoric-only-politicians} shows that they are widespread: 92.5\% of politicians in the survey sample experience at least one rhetoric-only day, with both blame-only and credit-only days represented. The design therefore relies on rhetoric-only days that are recurrent and widespread across politicians.

\subsection{Results for Voting Intentions}\label{sec:survey-result}

Blame increases voting support in the short run, while credit produces a smaller and imprecise effect. \Cref{tab:survey-main} reports estimates of Equation \ref{reg:survey} for all respondents in column (1). Support for the politician rises by 1 percentage point in the week after blame-only days, and the estimate is precise. The corresponding estimate after credit-only days is smaller, at 0.6 percentage points, and is not statistically significant. Blame therefore generates the clearer voting-intention response.

The dynamic pattern supports this timing interpretation. In \Cref{fig:survey_dynamic}, we estimate a dynamic version of Equation \ref{reg:survey} using three-day bins around rhetoric-only days to ensure sufficient precision. Support is flat in the two weeks before blame-only days: the pre-period coefficients are small and insignificant. After blame-only days, support increases in the first post-period bins and fades within two weeks. The corresponding estimates for credit-only days are smaller and show no systematic post-period increase. The support response therefore appears after blame-only days rather than reflecting pre-existing movement in support.

\begin{table}[t]
\caption{Voting Intentions}
\label{tab:survey-main}
\centering
\begin{footnotesize}
\begin{adjustbox}{max width=\textwidth}
\begin{tabular}{l*{3}{c}}
\toprule
& (1) & (2) & (3) \\
\midrule
& \multicolumn{1}{c}{All Respondents} & \multicolumn{2}{c}{Split by Social Media Use} \\
\cmidrule(lr){3-4}
& & No Social Media Use &  Social Media Use \\
\midrule
 & \multicolumn{3}{c}{(a) Blame}\\
\cmidrule(lr){2-4}
Post-Blame&0.010\sym{**}&0.006&0.014\sym{**}\\
&(0.005)&(0.010)&(0.006)\\
\midrule
District x Month FE  & \checkmark & \checkmark & \checkmark \\
Date FE  & \checkmark & \checkmark & \checkmark \\
Partisan Alignment FE  & \checkmark & \checkmark & \checkmark \\
Respondent Covariate FE  & \checkmark & \checkmark & \checkmark\\ 
\midrule
Observations &        20,154 &         5,781 &        14,300 \\
Clusters &           281 &           275 &           281 \\
\midrule
 & \multicolumn{3}{c}{(b) Credit} \\ 
\cmidrule(lr){2-4}
Post-Credit&0.006&-0.000&0.006\\
&(0.004)&(0.009)&(0.005)\\
\midrule
District x Month FE  & \checkmark & \checkmark & \checkmark \\
Date FE  & \checkmark & \checkmark & \checkmark \\
Partisan Alignment FE  & \checkmark & \checkmark & \checkmark \\
Respondent Covariate FE  & \checkmark & \checkmark & \checkmark\\ 
\midrule
Observations &        29,109 &         8,257 &        20,756 \\
Clusters &           325 &           322 &           325 \\
\bottomrule
\end{tabular}\end{adjustbox}
\\[1em]
\caption*{\scriptsize \textit{Notes}: The table presents estimates of Equation \ref{reg:survey}. The outcome is an indicator equal to one if the respondent reports intending to vote for the politician, and zero otherwise. Panel (a) reports estimates for blame-only days, and Panel (b) reports estimates for credit-only days. The treatment indicator equals one for respondents interviewed in the seven days after a blame-only or credit-only day. Column (1) uses all respondents. Column (2) restricts the sample to respondents who do not report having seen or heard political news on social media in the past week. Column (3) restricts the sample to respondents who do report having seen or heard political news on social media in the past week. All specifications include district-by-month, date, partisan alignment, and respondent covariate fixed effects. Standard errors in parentheses are clustered at the congressional-district level. *, **, *** denote significance at the 10\%, 5\%, and 1\% levels, respectively.\par}
\end{footnotesize}
\end{table}

\paragraph*{Social media channel}

The survey evidence also supports the social-media exposure channel. If the response operates through Twitter exposure to politicians' communication, it should be stronger among respondents more likely to encounter political news on social media. To test this prediction, we re-estimate Equation \ref{reg:survey} separately by whether respondents report having seen or heard news about politics on social media in the past week. \Cref{tab:survey-main} shows that the response is concentrated among these respondents in columns (2) and (3). Among social-media news users, support for the politician increases by 1.4 percentage points after blame-only days. Among respondents who do not report using social media for political news, the estimated effect is 0.4 percentage points and is not precisely estimated. The corresponding estimates for credit-only days are positive but imprecise. This heterogeneity reinforces the interpretation that the voting-intention response operates through exposure to politicians' communication on social media.

\paragraph*{Robustness checks} 

We estimate several additional specifications to assess the robustness of the survey results. These checks address alternative interpretations related to time-varying political shocks, Twitter activity around rhetoric-only days, and the definition of rhetoric-only days.

A first concern is that national or state-level events may shift both politicians' rhetoric and voter support differently across politicians' parties. Date fixed effects absorb shocks common to all districts, but they do not absorb shocks that differentially affect districts represented by Democratic and Republican politicians on the same day. For example, an event could lead Democratic politicians to use more blame while also increasing support for Democratic politicians for reasons unrelated to their own Twitter communication. To address this concern, we replace date fixed effects with date-by-politician-party fixed effects, which absorb daily shocks that differ by the politician's party. The results, reported in \Cref{tab:survey-date-robustness}, are essentially unchanged.

A second related concern is that events may shift support differently across respondent party groups while also changing the timing of blame-only or credit-only days. For example, an event could increase support among Democratic respondents and coincide with blame-only days by politicians in districts where Democratic respondents are more common. To address this concern, we replace date fixed effects with date-by-respondent-party-identification fixed effects, which absorb daily shocks that differ across respondent party groups. The results, also reported in \Cref{tab:survey-date-robustness}, are essentially unchanged.

A third concern is that rhetoric-only days may differ from nearby comparison days not only in rhetorical content, but also in the politician's Twitter activity on that day. If the timing of blame-only or credit-only days is correlated with unusual daily tweet volume, the estimated response could partly reflect that Twitter activity rather than the rhetorical style itself. To address this concern, we re-estimate Equation \ref{reg:survey} absorbing fixed effects for the number of tweets posted on the rhetoric-only day. \footnote{When multiple rhetoric-only days occur in the comparison window, we assign the maximum raw daily tweet volume among those days.} The results, reported in \Cref{tab:survey-volume-robustness}, are essentially unchanged.

A related concern is that rhetoric-only days may occur during periods of unusual communication activity more broadly, even if the rhetoric-only day itself is not unusual in volume. In that case, the estimates could reflect broader campaign intensity or political salience in the surrounding window, rather than voters' response to blame or credit rhetoric. To address this concern, we control for the log of total tweet volume across the full comparison window. The results, also reported in \Cref{tab:survey-volume-robustness}, are essentially unchanged.

We also test whether the results depend on our stringent definition of rhetoric-only days. The baseline specification defines a blame-only day as one on which all tweets posted by the politician are classified as blame, with credit-only days defined analogously. We re-estimate the survey specification using an alternative threshold based on the politician-specific 99th percentile of daily blame and credit intensity. This definition captures days on which a politician's communication is more concentrated in blame or credit than usual. The alternative definition is broader, but also potentially noisier, because daily rhetoric intensity is constructed from tweet-level binary indicators and can therefore be sensitive to small daily tweet counts. The results, reported in \Cref{tab:survey-p99}, are qualitatively unchanged across specifications, though somewhat attenuated. This attenuation is consistent with the broader and noisier definition, which includes less rhetorically clean communication and therefore weakens the contrast between treated and comparison periods.

\medskip
The survey evidence extends the donation results beyond selected donors and to a direct measure of intended vote choice. Support increases in the week after blame-only days, especially among respondents more likely to encounter political communication on social media, while credit-only days show no comparable response.

\subsection{Negative Rhetoric and the Returns to Blame}\label{sec:negativity}

The previous results show that blame generates positive political returns in both campaign contributions and voting intentions. A natural alternative explanation is that these returns reflect exposure to negative rhetoric rather than to blame in particular. However, our evidence does not support this interpretation. By definition, blame combines two dimensions of political rhetoric: it links political actions to outcomes and features a negative tone. If the returns to blame were simply driven by negativity, we would expect non-causal negative rhetoric, that is, rhetoric with a negative tone that does not link political actions to outcomes, to generate positive political returns as well. To assess this possibility, we apply the same empirical designs to non-causal negative and positive rhetoric.

We first return to campaign contributions. We re-estimate our baseline donation specification replacing blame and credit with the shares of non-causal negative and non-causal positive tweets, while otherwise keeping the empirical design unchanged. The results are reported in Panel (a) of \Cref{tab:negativity}. Non-causal negative rhetoric does not reproduce the returns to blame. The estimated effect on donation revenue is small and negative. Thus, applying the same exposure design to non-causal negative rhetoric does not generate the positive returns we observe for blame.

We then turn to voting intentions. We re-estimate our event-study specification using non-causal-negative-only and non-causal-positive-only days in place of blame-only and credit-only days, while otherwise keeping the empirical design unchanged. The results are reported in Panel (b) of \Cref{tab:negativity}. We find no statistically significant increase in support following non-causal-negative-only days. Thus, applying the same event-study design to non-causal negative rhetoric does not generate the statistically significant positive voting-intention response we observe for blame.

\medskip
Taken together, the evidence in this section shows that the political returns to causal rhetoric are concentrated in blame. Blame generates positive returns in both campaign contributions and voting intentions, while credit shows no comparable aggregate pattern. We find no evidence that the returns to blame are simply driven by exposure to negativity: non-causal negative rhetoric does not reproduce the consistently positive pattern observed for blame across the two settings. Having established these returns, we next ask how well grounded the causal accounts politicians provide are in economic knowledge.

\section{The Economic Grounding of Causal Rhetoric}\label{sec:content}

The preceding sections show that causal rhetoric is widely used in congressional communication and that its political returns are concentrated in blame. We next ask a further question about the accounts politicians supply: how well grounded are the causal accounts politicians provide? Because these accounts are what politicians are trying to convince voters of, it matters whether voters are being convinced of claims that experts would support and agree with, or of claims that require substantial qualification. We study this question in the economic domain, where causal rhetoric is common and where claims linking political actions to outcomes can be assessed against standard macroeconomic reasoning.

We evaluate economic causal claims based on whether they are supported as stated. Determining whether these claims are simply true or false would be difficult because their validity often depends on assumptions about timing, magnitude, incidence, behavioral responses, or general-equilibrium effects. We therefore do not attempt to classify claims as true or false. Instead, we ask how much qualification is needed for a claim, as expressed in the tweet, to be supported by standard macroeconomic reasoning. We find that most economic causal claims are not supported as stated and require qualifications that the tweet does not provide. At the same time, outright unsupported claims are rare. The need for qualification is substantially greater for blame than for credit.

\subsection{Assessing Economic Claims as Stated}

We analyze 413,836 tweets that are both classified as blame or credit and identified by our topic classifier as being about the economy. For each tweet, we use an AI-assisted coding procedure to extract the dominant causal claim, identify the implied economic mechanism, and classify the claim's level of support.\footnote{We implement this procedure using GPT-5 mini under a fixed codebook and prompt that instruct the model to act as a careful research coder and evaluate claims using standard macroeconomic reasoning. Appendix \ref{app:assessing} provides additional details on the prompt, codebook, and implementation.}

The classification assigns each claim to one of four categories. A claim is \emph{supported} if standard macroeconomic reasoning supports it under ordinary assumptions, without requiring important additional qualifications. A claim is \emph{minor qualified} if the basic mechanism and direction are standard, but the claim requires familiar caveats about context, timing, magnitude, or incidence. A claim is \emph{major qualified} if some route to support exists, but only after introducing substantial unstated conditions or offsetting forces. A claim is \emph{unsupported} if no ordinary macroeconomic route supports the claim under its baseline interpretation.

The resulting measure captures the degree of qualification needed to reconcile the claim as stated with standard macroeconomic reasoning. It is therefore deliberately different from a binary measure of whether a claim is true or false. A qualified claim may rely on a standard economic mechanism, but state the relationship too categorically given the conditions under which that mechanism operates. Conversely, the unsupported category is reserved for claims for which no ordinary route to support exists under the baseline interpretation.

\subsection{Economic Grounding of Blame and Credit}

Most economic causal claims are not supported as stated but require qualification. \Cref{fig:macroexpert_all} reports the distribution of support labels across economic blame and credit tweets. Only about 10 percent of claims are classified as supported as stated. Roughly 90 percent require either minor or major qualifications. The largest category is minor qualified, at 62 percent, followed by major qualified, at 28 percent. Unsupported claims are rare, accounting for less than 1 percent of observations.

\begin{figure}[t]
    \centering
    \caption{Support Level for Economic Causal Rhetoric}
    \label{fig:macroexpert}
    \begin{subfigure}{0.49\textwidth}
    \caption{Pooled}
    \label{fig:macroexpert_all}
    \includegraphics[width=\linewidth]{inputs_paper/current/figures/macroexpert_all.pdf}
    \end{subfigure}
    \begin{subfigure}{0.49\textwidth}
    \caption{Split by Blame and Credit}
    \label{fig:macroexpert_split}
    \includegraphics[width=\linewidth]{inputs_paper/current/figures/macroexpert_split.pdf}
    \end{subfigure}
    \vspace{0.3cm}
    \caption*{\scriptsize \textit{Notes}: Panel (a) reports the distribution of macroeconomic-support labels across tweets classified as blame or credit and identified by our topic classification as being about the economy. Panel (b) reports the same distribution separately for tweets classified as blame and credit.}
\end{figure}

The need for qualification is especially pronounced for blame. \Cref{fig:macroexpert_split} reports the same classification separately for blame and credit tweets. Blame claims are much more likely to be major qualified: 44 percent of blame claims fall in this category, compared with 16 percent of credit claims. Credit claims are instead more concentrated in the minor qualified category, which accounts for 75 percent of credit claims and 46 percent of blame claims. Unsupported claims remain rare for both styles, but they are more common among blame tweets: 1.8 percent of blame claims are unsupported, compared with 0.17 percent of credit claims.

\medskip
Taken together, the evidence shows that economic causal claims are rarely supported as stated. Most require qualifications that the tweet does not provide, while outright unsupported claims are rare. The need for qualification is substantially greater for blame than for credit. Combined with the evidence in the previous section, this reveals a sharp asymmetry: blame is both the rhetorical style with the largest political returns and the one whose economic causal claims most often require substantial qualification.

\section{Additional Patterns}\label{sec:additional}

The preceding sections establish the paper's main findings on the supply and political returns of causal rhetoric and further assess the economic grounding of the claims politicians supply. We close the empirical analysis by documenting two additional patterns in how blame and credit are transmitted and deployed. First, we examine amplification on social media. Blame receives substantially more retweets than credit or other rhetoric, indicating that it travels farther on the platform and providing a possible amplification channel consistent with its larger political returns. Second, we examine whether politicians use blame and credit in politically convenient ways. In the economic domain, blame becomes more common when the opposing party is legislatively active, while credit becomes more common when one's own party is active. Together, these exercises provide additional insight into how causal rhetoric is transmitted and deployed in politicians' communication.

\subsection{Blame and Amplification}\label{sec:amplification}

The returns evidence shows that blame generates the largest gains in political support. One possible reason is amplification: if blame travels farther on social media, more users may encounter it. We therefore examine whether blame is disproportionately associated with retweet activity.

Retweets provide a direct and informative measure of how widely a tweet diffuses on the platform and therefore of how far it travels. This makes retweet activity the relevant outcome for studying amplification. Retweet counts are available through March 2020 and cover roughly half of the tweets in our corpus over that period. \Cref{tab:virality-attrition} shows that this engagement subsample is broadly representative of the underlying tweet sample, reducing concerns that the results are driven by selection into non-missing retweet data. \Cref{app:virality} provides additional details on the data and empirical specifications.

Blame receives disproportionate amplification. In the sample with non-missing retweet data, blame tweets account for only 15\% of tweets but nearly 40\% of all retweets. \Cref{fig:virality_hist} shows that blame tweets receive, on average, nearly four times as many retweets as tweets in the other rhetorical categories. Conditional on being posted, blame is therefore much more likely to circulate widely.

\begin{figure}
    \centering
    \caption{Blame and Retweet Amplification}
    \label{fig:virality_hist}
    \includegraphics[width=0.5\linewidth]{inputs_paper/current/figures/virality_hist.pdf}
    \caption*{\scriptsize \textit{Notes}: The figure reports the average number of retweets received by tweets classified as blame, credit, and neither blame nor credit, restricting to tweets with non-missing retweet data. Bars represent category means and intervals represent 95 percent confidence intervals.}
\end{figure}

More structured analyses show the same pattern. \Cref{fig:virality_static} shows that, relative to tweets not classified as blame or credit, blame tweets receive about 0.15 standard deviations more retweets after controlling for politician-by-date and topic-by-date fixed effects. Credit tweets receive no comparable retweet premium. \Cref{fig:virality_quantile} shows that this relationship is strongest in the upper tail of the retweet distribution: blame is negatively associated with appearing in the lower deciles of retweet counts and increasingly positively associated with appearing in the upper deciles, with the largest estimate in the top decile.

These patterns are consistent with amplification contributing to blame's larger political returns. The rhetorical style with the largest returns also travels substantially farther on social media. This evidence connects to \citet{boken2026returns}, who show that small donations rise sharply when politicians go viral.

\subsection{Politically Convenient Deployment of Causal Rhetoric}\label{sec:bills}

We next ask whether politicians deploy blame and credit in politically convenient ways. Because blame assigns responsibility for negative outcomes while credit links political actors to positive outcomes, the political usefulness of the two styles should depend on which party can be associated with the relevant action. When the opposing party is active, blame is politically convenient; when one's own party is active, credit is politically convenient. We examine whether politicians' use of economic causal rhetoric follows this pattern around congressional legislative activity.

We construct a dataset of congressional bills using official records from the U.S. Government Publishing Office, accessed through the GovInfo bulk data repository. We collect structured metadata for bills introduced in the U.S. House and Senate from the 112th Congress (2011--2013) through the 118th Congress (2023--2025).\footnote{Specifically, we include House bills, House joint resolutions, House concurrent resolutions, and House simple resolutions, as well as their Senate counterparts.} To focus on politically salient legislation, we restrict attention to bills sponsored by Democratic or Republican legislators that received at least one roll-call vote. We further restrict the sample to the period in which Senators are observed in our Twitter data and to bills whose metadata identify them as belonging to an economic policy area.\footnote{We identify economic bills using the policy-area field included in the GovInfo metadata. We keep bills in the following policy areas: Agriculture and Food, Commerce, Economics and Public Finance, Energy, Finance and Financial Sector, Foreign Trade and International Finance, Housing and Community Development, Labor and Employment, Social Welfare, Taxation, and Transportation and Public Works.} The resulting sample contains 327 economic bills.

We construct indicators of own-party and opposing-party legislative activity at the date-by-chamber-by-party level, separately for the House and Senate. For each politician-day, we define $\text{OwnBill}_{it}$ as an indicator equal to one if, on date $t$, at least one economic bill is introduced in politician $i$'s chamber by a member of politician $i$'s party. We define $\text{OppBill}_{it}$ analogously for economic bills introduced in the same chamber by members of the opposing party.

We restrict the analysis to the period in which both House and Senate tweets are observed and include all 537,998 tweets identified by our topic classifier as being about the economy, whether or not they contain causal rhetoric. At the tweet level, let $y_{\ell it}$ be an indicator equal to one if economic tweet $\ell$, posted by politician $i$ on date $t$, is classified as rhetorical style $y$, where $y \in \{\text{Blame}, \text{Credit}\}$. We estimate separate specifications for blame and credit:
\begin{align}
    y_{\ell it}
    = \beta_{\text{Opp}} \text{OppBill}_{it}
    + \beta_{\text{Own}} \text{OwnBill}_{it}
    + \lambda_{i \times m(t)}
    + \mu_{t \times P_i}
    + \varepsilon_{\ell it}.
    \label{reg:bills}
\end{align}
The fixed effects $\lambda_{i \times m(t)}$ are politician-by-month fixed effects, which absorb politician-specific differences in rhetorical style and month-specific changes in each politician's communication. The fixed effects $\mu_{t \times P_i}$ are date-by-party fixed effects, where $P_i$ denotes politician $i$'s party, and absorb shocks common to politicians from the same party on the same date. Standard errors are clustered by date.

The coefficients describe conditional differences in the use of blame and credit around own-party and opposing-party economic legislative activity. Date-by-party fixed effects compare politicians from the same party on the same date, while politician-by-month fixed effects compare each politician with themselves within the same month. The remaining variation comes from differences in economic bill introductions across the House and Senate within the same party and date. Because bill introductions are political events rather than randomly assigned shocks, we interpret the estimates descriptively.

\begin{table}[t]
\caption{Economic Causal Rhetoric around Economic Bill Introductions}
\label{tab:bills-chambers-econ}
\centering
\begin{footnotesize}
\begin{adjustbox}{max width=\textwidth}
\begin{tabular}{l*{2}{c}}
\toprule
& (1) & (2) \\
\midrule
& \multicolumn{1}{c}{Blame} & \multicolumn{1}{c}{Credit} \\
\midrule
Bill introduced by opposing party&0.014\sym{**}&-0.002\\
&(0.007)&(0.006)\\
Bill introduced by own party&-0.006&0.018\sym{**}\\
&(0.006)&(0.007)\\
\midrule
Politician x Month FE  & \checkmark & \checkmark \\
Date x Party FE  & \checkmark & \checkmark \\
\midrule
Observations &       535,776 &       535,776 \\
Clusters &         2,203 &         2,203 \\
\bottomrule
\end{tabular}\end{adjustbox}
\\[1em]
\caption*{\scriptsize \textit{Notes}: The table presents estimates of Equation \ref{reg:bills}. The sample includes all tweets classified as being about the economy. The outcome in column (1) is an indicator equal to one if the tweet is classified as blame. The outcome in column (2) is an indicator equal to one if the tweet is classified as credit. The regressors indicate whether, on the date of the tweet, at least one economic bill was introduced in the politician's chamber by a member of the opposing party or by a member of the politician's own party. All specifications include politician-by-month and date-by-politician-party fixed effects. Standard errors in parentheses are clustered by date. *, **, *** denote significance at the 10\%, 5\%, and 1\% levels, respectively.\par}
\end{footnotesize}
\end{table}

\Cref{tab:bills-chambers-econ} shows a clear pattern of politically convenient deployment. On dates when the opposing party introduces an economic bill in a politician's chamber, economic tweets by politicians in that chamber are 1.4 percentage points more likely to be classified as blame, relative to economic tweets by same-party politicians on the same date. Own-party economic bill introductions are not associated with a comparable increase in blame. For credit, the pattern is reversed: own-party economic bill introductions are associated with a 1.8 percentage point higher probability that an economic tweet is classified as credit, while opposing-party economic bill introductions are not associated with a comparable increase.

Politicians therefore use the two forms of causal rhetoric in systematically different political circumstances. Blame is more common when the opposing party is legislatively active, whereas credit is more common when one's own party is active. This pattern is politically convenient: politicians assign blame when political action can be associated with the opposing party and claim credit when it can be associated with their own party. Causal rhetoric is therefore deployed in a politically convenient way.

\medskip
Taken together, these additional patterns characterize how causal rhetoric is transmitted and deployed. Blame receives substantially greater amplification on social media by attracting more retweets, a pattern consistent with its larger political returns. Blame and credit are also used in politically convenient ways, with blame associated with opposing-party legislative activity and credit with own-party activity. This section complements the paper's main evidence by providing further insights into the amplification and deployment of causal rhetoric.

\section{Conclusion}\label{sec:conclusion}

Democratic politics asks voters to form beliefs about how political actions map into outcomes. This paper studies the supply and political returns of a rhetorical margin in that process: politicians' use of language that links the actual or potential action of a political actor to an outcome through blame and credit. We call this language causal rhetoric and measure its use in congressional tweets by fine-tuning a language model. Causal rhetoric is a central part of political communication. It accounts for more than one third of congressional tweets, is the language of political disagreement, and has become substantially more common over time.

The political returns to causal rhetoric are asymmetric and concentrated in blame. Exposure to blame generates sizable gains in political support, while credit produces no comparable aggregate response. In campaign-contributions data, blame increases small-donation revenue primarily by mobilizing additional donors, while credit produces smaller returns detectable only in average giving. Survey evidence for the 2020 House elections points in the same direction: support rises after blame-only days, especially among respondents who use social media for political news. We also find no evidence that these returns simply reflect exposure to negativity: non-causal negative rhetoric does not reproduce the significant positive pattern observed for blame across the two settings

We also assess how well grounded the causal accounts politicians supply are by examining economic claims. Most are not supported as stated and require qualifications that the tweet does not provide, while outright unsupported claims are rare. These qualifications are substantially more important for blame than for credit, making blame the least economically grounded form of causal rhetoric.

Finally, two additional patterns further characterize how causal rhetoric is transmitted and deployed. Blame receives greater amplification on social media through retweets, while politicians use blame and credit in politically convenient ways around economic legislative activity.

Overall, the paper shows that causal rhetoric is a major input into democratic politics. Politicians frequently tell voters how political actions and outcomes are connected, and these claims generate political returns, especially when they assign blame. At the same time, causal rhetoric, especially blame, is rarely supported by economic knowledge. The broader implication is that fully understanding accountability in democratic politics also requires studying how politicians can shape the causal accounts through which voters evaluate them.

\FloatBarrier
\newpage
\bibliographystyle{chicago}
\bibliography{references.bib}

\newpage
\appendix

\renewcommand{\thefigure}{\Alph{section}\arabic{figure}}
\renewcommand{\thetable}{\Alph{section}\arabic{table}}
\renewcommand{\thelstlisting}{\Alph{section}\arabic{lstlisting}}

\makeatletter
\@addtoreset{figure}{section}
\@addtoreset{table}{section}
\@addtoreset{lstlisting}{section}
\makeatother

\section{Additional Figures and Tables}\label{app:additional}

\subsection{Data and Measurement}\label{app:additional_data_measurement}

\begin{table}[H]
\caption{Descriptive Statistics}
\label{tab:descriptives}
\centering
\begin{footnotesize}
\begin{adjustbox}{max width=\textwidth}
\begin{tabular}{lcccccc}
\toprule
& \multicolumn{3}{c}{(a) Tweet Level} & \multicolumn{3}{c}{(b) Politician Level} \\
\cmidrule(lr){2-4} \cmidrule(lr){5-7}
& All & Democrats & Republicans & All & Democrats  & Republicans\\
\midrule
Female&0.300&0.401&0.165&0.243&0.378&0.132\\
&(0.458)&(0.490)&(0.371)&(0.429)&(0.486)&(0.339)\\
Age&57.973&59.058&56.509&57.468&58.499&56.617\\
&(11.572)&(12.075)&(10.682)&(11.767)&(12.649)&(10.926)\\
Black&0.082&0.127&0.021&0.080&0.150&0.022\\
&(0.274)&(0.333)&(0.142)&(0.271)&(0.357)&(0.148)\\
Bachelor&0.333&0.297&0.381&0.332&0.295&0.363\\
&(0.471)&(0.457)&(0.486)&(0.471)&(0.457)&(0.481)\\
Master or Higher&0.620&0.672&0.551&0.584&0.651&0.529\\
&(0.485)&(0.469)&(0.497)&(0.493)&(0.477)&(0.500)\\
Republican&0.426&0&1&0.548&0&1\\
&(0.494)&(0)&(0)&(0.498)&(0)&(0)\\
Nominate Score&-0.005&-0.387&0.511&0.109&-0.366&0.500\\
&(0.467)&(0.126)&(0.164)&(0.455)&(0.122)&(0.162)\\
Representative&0.804&0.814&0.790&0.857&0.857&0.856\\
&(0.397)&(0.389)&(0.407)&(0.351)&(0.350)&(0.351)\\
Senator&&&&0.118&0.123&0.114\\
&&&&(0.323)&(0.329)&(0.318)\\
Both Representative and Senator&&&&0.026&0.020&0.030\\
&&&&(0.158)&(0.139)&(0.172)\\
Number of Tweets Posted&&&&4664.950&5924.391&3625.209\\
&&&&(4848.342)&(5151.747)&(4319.150)\\
Number of Accounts&&&&1.988&2.042&1.943\\
&&&&(0.689)&(0.688)&(0.687)\\
&&&&&&\\
Observations&4,198,455&2,411,227&1,787,228&900&407&493\\
\bottomrule
\end{tabular}\end{adjustbox}
\\[1em]
{\raggedright \scriptsize \textit{Notes}: Panel (a) presents statitics at the tweet level. Panel (b) presents statistics at the politician level. Means and standard deviations in parentheses.\par}
\end{footnotesize}
\end{table}

\begin{table}[H]
\caption{Descriptive Statistics of Donations}
\label{tab:descriptives_donations}
\centering
\begin{footnotesize}
\begin{adjustbox}{max width=\textwidth}
\begin{tabular}{lcc}
\toprule
& Donation Level & Donor Level \\
\midrule
Mean&36.40&318.62\\
10th percentile&3.00&8.75\\
25th percentile&5.00&23.00\\
50th percentile&15.00&53.93\\
75th percentile&27.00&211.00\\
90th percentile&95.00&612.51\\
&&\\
Observations&40,087,964&4,580,004\\
\bottomrule
\end{tabular}
\end{adjustbox}
\\[1em]
{\raggedright \scriptsize \textit{Notes}: The table reports the mean and the 10th, 25th, 50th, 75th, and 90th percentiles of amount. All amounts are in U.S. dollars. At the donation level, amount is the value of an individual donation. At the donor level, amount is total giving by the donor in our data.\par}
\end{footnotesize}
\end{table}

\begin{figure}[H]
    \centering
    \caption{Distributions}
    \label{fig:p_dists}
    \begin{subfigure}{0.32\textwidth}
        \caption{Blame}
        \label{fig:p_dists_blame}
        \includegraphics[width=\linewidth]{inputs_paper/current/figures/density_p_blame.pdf}
    \end{subfigure}
    \begin{subfigure}{0.32\textwidth}
        \caption{Credit}
        \label{fig:p_dists_merit}
        \includegraphics[width=\linewidth]{inputs_paper/current/figures/density_p_merit.pdf}
    \end{subfigure}
    \begin{subfigure}{0.32\textwidth}
        \caption{None}
        \label{fig:p_dists_none}
        \includegraphics[width=\linewidth]{inputs_paper/current/figures/density_p_none.pdf}
    \end{subfigure}
    \vspace{0.3cm}
    \caption*{\scriptsize \textit{Notes}: Each panel reports the distribution of the classifier-assigned probability for the predicted class, conditional on that class being selected. Panel (a) shows the distribution of $P(\text{Blame})$ among tweets classified as blame, that is, tweets for which $P(\text{Blame}) > \max\{P(\text{Credit}), P(\text{None})\}$. Panels (b) and (c) report the analogous distributions for credit and none.}
\end{figure}

\begin{table}[H]\centering
    \def\sym#1{\ifmmode^{#1}\else\(^{#1}\)\fi}
    \caption{Labeled Tweets}
    \label{tab:sample_tweets_labeled}
    \begin{adjustbox}{max width=\textwidth}
    \footnotesize
    \begin{tabular}{
        >{\raggedright\arraybackslash}p{0.7\linewidth} 
        >{\centering\arraybackslash}p{0.06\linewidth} 
        >{\centering\arraybackslash}p{0.06\linewidth} 
        >{\centering\arraybackslash}p{0.15\linewidth}
    }
    \toprule
    Tweet & Causal & Tone & Rhetoric Style \\
    \hline
    \rowcolor{gray!10}
    \scriptsize{Biden has lost all credibility. No one believes your lies, Joe! URL } & 0 & -1 & None\\
    \scriptsize{.\@realDonaldTrump is a terrible role model for our kids. They deserve better. \#Debate} & 0 & -1 & None\\
    \rowcolor{gray!10}
    \scriptsize{Joe Biden and the Democrats’ terrible policies have wreaked havoc on this country.} & 1 & -1 & Blame\\
    \scriptsize{\#Trumpcare is fundamentally flawed. Higher costs, less coverage, fewer protections -- that’s GOP’s gift to American people. \#ProtectOurCare URL} & 1 & -1 & Blame\\
    \rowcolor{gray!10}
    \scriptsize{I couldn’t support final passage of today’s approps package but I’m pleased about the inclusion of my HBCU amendment URL} & 0 & 1 & None\\
    \scriptsize{Universal congrats to the scientists at @OregonState for their work helping Insight make a successful \#MarsLanding: URL} & 0 & 1 & None\\
    \rowcolor{gray!10}
    \scriptsize{The Protect Medical Innovation Act will boost American innovation and manufacturing, and it will encourage medical research and development that make a real difference in people’s lives. URL} & 1 & 1 & Credit\\
    \scriptsize{Now that the Inflation Reduction Act is law it will not only lower prescription drug prices but save lives. Thank you @HenryFordHealth for your support. URL} & 1 & 1 & Credit\\
    \bottomrule
    \end{tabular}
    \end{adjustbox}
    \\[0.5em]
    {\raggedright \scriptsize \textit{Notes}: The table reports illustrative tweets from the hand-labeled dataset used to train the classifier. For each tweet, we report the manual labels for causality and tone, together with the implied rhetorical-style classification.\par}
\end{table}

\begin{figure}[H]
    \centering
    \caption{Linguistic Features}
    \label{fig:linguistic_features}
    \begin{subfigure}{0.32\textwidth}
        \caption{\scriptsize Self- versus Other-Referential}
        \label{fig:linguistic_features_selfother}
        \includegraphics[width=\linewidth]{inputs_paper/current/figures/linguistic_features_selfother.pdf}
    \end{subfigure}
    \begin{subfigure}{0.32\textwidth}
        \caption{\scriptsize Attributions to the Opposing Party}
        \label{fig:linguistic_features_entities}
        \includegraphics[width=\linewidth]{inputs_paper/current/figures/linguistic_features_entities.pdf}
    \end{subfigure}
    \begin{subfigure}{0.32\textwidth}
        \caption{\scriptsize Future versus Past Orientation}
        \label{fig:linguistic_features_tense}
        \includegraphics[width=\linewidth]{inputs_paper/current/figures/linguistic_features_tense.pdf}
    \end{subfigure}
    \vspace{0.3cm}
    \caption*{\scriptsize \textit{Notes}: The figure reports linguistic and semantic measures across rhetorical styles. Panel (a) compares the relative prevalence of second- and third-person pronouns to first-person pronouns, capturing whether tweets are more other- or self-referential. Panel (b) reports the share of tweets that target the opposing party, among tweets for which party identity can be identified. Panel (c) compares the use of future and past tense, capturing whether the rhetoric is more prospective or retrospective. The measures in panels (a) and (c) are standardized. Bars represent 95 percent confidence intervals.}
\end{figure}

\begin{figure}[H]
    \centering
    \caption{Emotions}
    \label{fig:emotions}
    \includegraphics[width=0.8\linewidth]{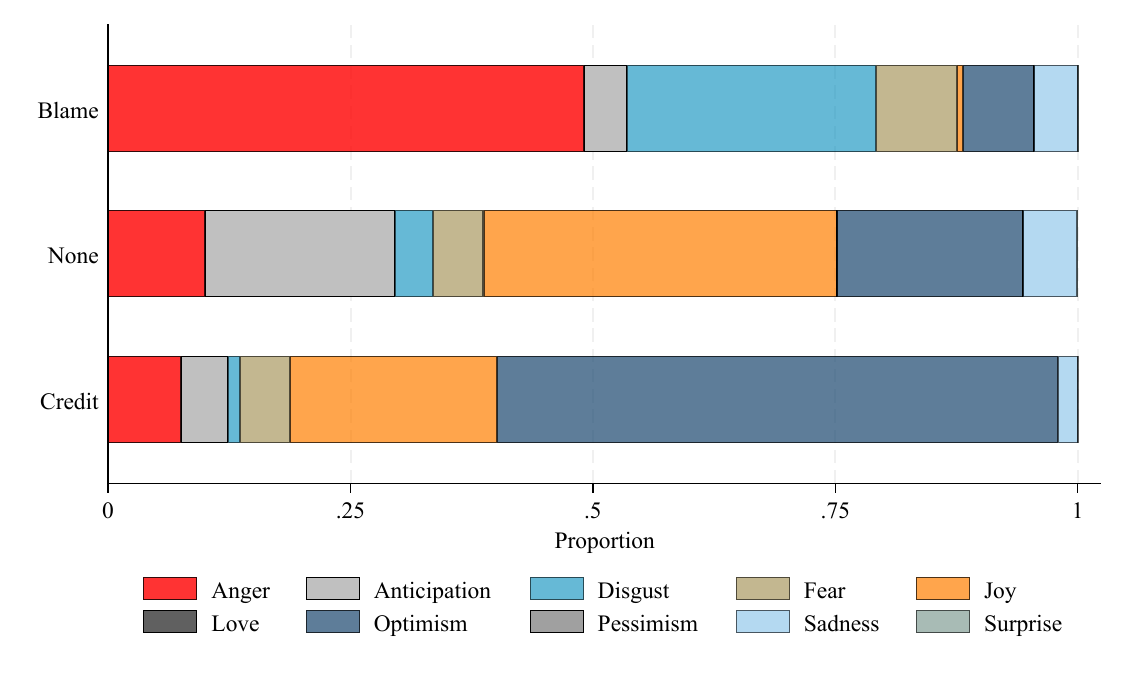}
    \caption*{\scriptsize \textit{Notes}: Each bar reports the share of tweets within a given rhetorical style classified into the corresponding emotion category.}
\end{figure}

\begin{figure}[H]
    \centering
    \caption{Bigrams Diagnostic of Blame and Credit for Democrats}
    \label{fig:distinctiveness_bigrams_dems}
    \includegraphics[width=\linewidth]{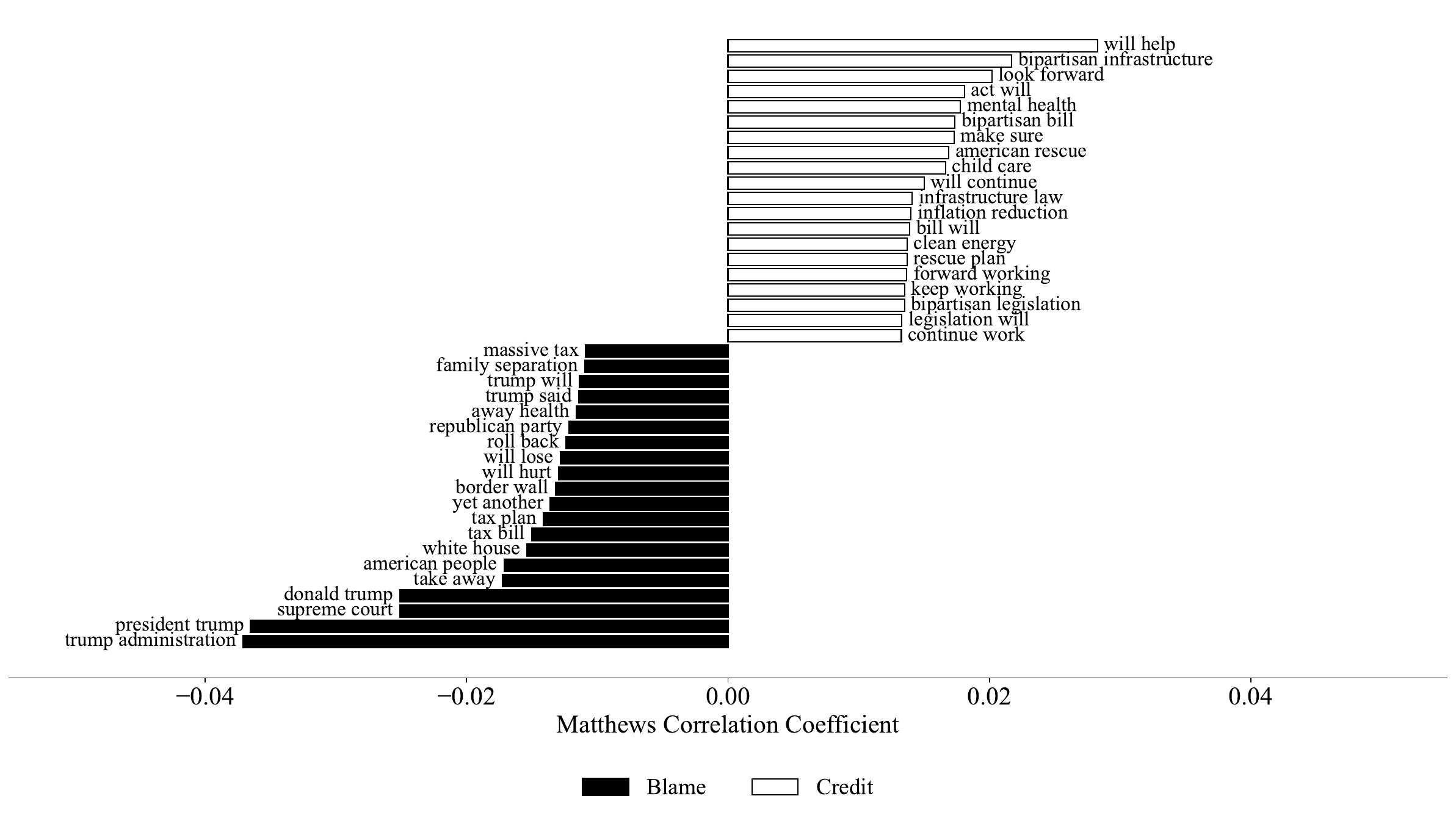}
    \caption*{\scriptsize \textit{Notes}: The figure reports the 20 bigrams that are most distinctive of blame and credit tweets among Democratic politicians, ranked by their Matthews correlation coefficient.}
\end{figure}

\begin{figure}[H]
    \centering
    \caption{Bigrams Diagnostic of Blame and Credit for Republicans}
    \label{fig:distinctiveness_bigrams_reps}
    \includegraphics[width=\linewidth]{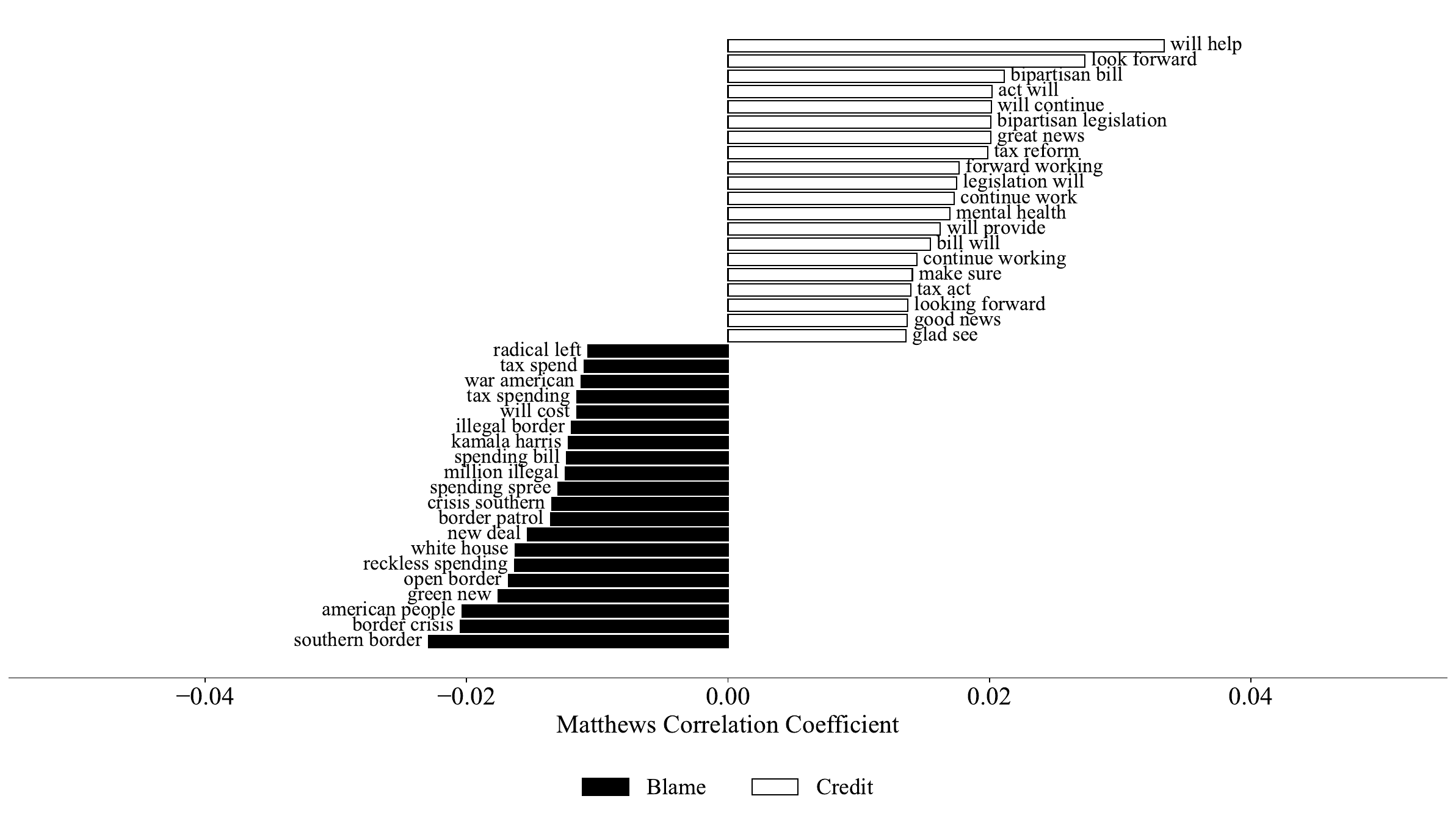}
    \caption*{\scriptsize \textit{Notes}: The figure reports the 20 bigrams that are most distinctive of blame and credit tweets among Republican politicians, ranked by their Matthews correlation coefficient.}
\end{figure}

\begin{table}[H]\centering
    \def\sym#1{\ifmmode^{#1}\else\(^{#1}\)\fi}
    \caption{Democrats Tweets Classified as Blame and Credit}
    \label{tab:sample_tweets_dems}
    \begin{adjustbox}{max width=\textwidth}
    \scriptsize
    \begin{tabular}{p{\linewidth}}
    \toprule
    (a) Blame Tweets\\
    \hline 
    \rowcolor{gray!10}
    We warned when the GOP passed tax cuts for the rich that it would explode deficits. It did. We warned that the GOP would use those deficits to come after Social Security and Medicare. They are. URL\\
    The party of NO, \#ILGOP in particular, plays political games in ignoring the implications on our economy, on jobs, Social Security checks. Republicans raised the debt ceiling 3 times under Trump's thumb. They are playing politics with people's lives. URL\\
    \rowcolor{gray!10}
    GOP’s reckless health care strategy is already destabilizing \#healthcare markets and forcing premiums to rise.\\
    If \#GOP could govern, we wouldn't be having this conversation. The root of \#TrumpShutdown is incompetence, putting veterans and kids health care at risk, pensions at risk, and the fight against opioids at risk\\
    \rowcolor{gray!10}
    Real wages today are lower than they were in 1973. That’s not a sign of a healthy economy, it’s a sign that working people today are worse off than they were 45 years ago, and the GOP tax cuts have done nothing to address that issue.\\
    \midrule
    (b) Credit Tweets\\
    \hline
    \rowcolor{gray!10}
    @EPAScottPruitt \#CleanAirAct has both strengthened our economy and protected our health. Dangerous air pollutants have been cut by 68\%, while GDP grew 212\%.\\
    8 million private sector jobs have been added since the \#ACA has been implemented. URL\\
    \rowcolor{gray!10}
    The Bipartisan Infrastructure Law is putting our economy on track to thrive \& investing in communities that have too often been left behind. With over \$2 million recently announced, we take a major step toward redeveloping Baltimore’s ‘Bridge to Nowhere.’ URL\\
    Happy to be joining @HouseDemocrats to help America’s workers access better paying jobs. The Workforce Innovation and Opportunity Act connects employers with qualified candidates, lowers costs for families and increases supplies. Democrats are \#InvestingInWorkers. URL\\
    \rowcolor{gray!10}
    The \#BuildBackBetterAct provides much needed funds to @TheJusticeDept to help reduce community violence \& fund proven intervention programs. I’m proud to advocate for legislation to break cycles of violence in communities, saving American lives \& taxpayer dollars.\\
    \bottomrule
    \end{tabular}
    \end{adjustbox}
    \\[0.5em]
    {\raggedright \scriptsize \textit{Notes}: The table reports illustrative examples of tweets classified as blame and credit among Democratic politicians. Panel (a) reports blame tweets and Panel (b) reports credit tweets. These examples are intended to provide qualitative intuition for the rhetorical patterns captured by our classification.\par}
\end{table}
\begin{table}[H]\centering
    \def\sym#1{\ifmmode^{#1}\else\(^{#1}\)\fi}
    \caption{Republicans Tweets Classified as Blame and Credit}
    \label{tab:sample_tweets_reps}
    \begin{adjustbox}{max width=\textwidth}
    \scriptsize
    \begin{tabular}{p{\linewidth}}
    \toprule
    (a) Blame Tweets\\
    \hline 
    \rowcolor{gray!10}
    The supply chain and inflation crises are not a “high class problem” like @WHCOS claims. As Dems look to pour trillions into the economy and spike inflation further, they must understand that actions have consequences that will be felt by every American URL\\
    Our country is facing soaring inflation thanks to Democrats’ spending spree, and what’s @POTUS’ solution? Spend MORE money.\\
    \rowcolor{gray!10}
    @JoeBiden and @SenateDems are TOTALLY out of touch with reality. Inflation is still wiping out wage growth, all while Democrats' reckless spending spree makes matters worse. URL\\
    Top border officials told Biden that if he unraveled Trump's policies and pushed for open borders that a major crisis would occur. He didn’t listen. Now everyone is suffering -- Americans and migrants alike. URL\\
    \rowcolor{gray!10}
    April saw the highest number of migrants ever recorded. Next week, @JoeBiden will reverse another commonsense border policy that will only make this crisis worse. Biden needs to wake up and face reality. URL\\
    \midrule
    (b) Credit Tweets\\
    \hline
    \rowcolor{gray!10}
    According to the @JECRepublicans, 1 million jobs have been created since tax reform was enacted. Learn more about \#taxreform at URL or text ``tax reform'' to 50589. URL\\
    Today @realDonaldTrump showed his commitment to supporting American energy dominance. The @EPA's rule will bolster our nation’s energy independence by lowering energy costs, spurring job growth and promoting economic development in our communities. URL\\
    \rowcolor{gray!10}
    Glad to hear @realDonaldTrump has signed legislation adding \$320 billion to the \#PaycheckProtectionProgram, ramping up testing capability and providing more funding for health care providers. AR will benefit from this measure to protect public health and save businesses \& jobs. URL URL\\
    Earlier this week, I introduced \#CARA2 to increase funding levels for programs we know work and implements additional policy reforms that will make a real difference in combatting the \#opioidcrisis. URL\\
    \rowcolor{gray!10}
    @TransportGOP are delivering on our promise of fixing supply chain holes and building a stronger economy. Currently we are marking up a package of bills that will remove barriers, increase efficiency, and target infrastructure investment. \#SupplyChain\\
    \bottomrule
    \end{tabular}
    \end{adjustbox}
    \\[0.5em]
    {\raggedright \scriptsize \textit{Notes}: The table reports illustrative examples of tweets classified as blame and credit among Republican politicians. Panel (a) reports blame tweets and Panel (b) reports credit tweets. These examples are intended to provide qualitative intuition for the rhetorical patterns captured by our classification.\par}
\end{table}

\begin{figure}[H]
    \centering
    \caption{Distinctiveness Relative to Existing Text Measures}
    \label{fig:coefs_text}
    \includegraphics[width=0.5\linewidth]{inputs_paper/current/figures/coefs_text.pdf}
    \caption*{\scriptsize \textit{Notes}: The figure reports coefficients from tweet-level regressions in which the outcome is an indicator equal to one if the tweet is classified as blame or credit. Each outcome is regressed on three standardized text measures: sentiment, emotionality, and moral terminology. Sentiment is measured using the \citet{camacho2022tweetnlp} model, which assigns each tweet probabilities of being negative, neutral, or positive; we convert these probabilities into a continuous sentiment measure equal to $-1 \times \Pr(\text{negative}) + 0 \times \Pr(\text{neutral}) + 1 \times \Pr(\text{positive})$. Emotionality follows \citet{gennaro2022emotion}: we embed emotional and reasoning words and compute, for each tweet, the relative cosine similarity of the tweet embedding to the emotional-word embedding versus the reasoning-word embedding. Moral terminology follows \citet{enke2020moral}: for each moral value, we compute the average frequency of vice and virtue terms, sum across moral values, and normalize by the number of non-stop words. All regressors are standardized. Bars represent 95 percent confidence intervals computed with standard errors clustered at the politician level.\par}
\end{figure}

\FloatBarrier
\newpage
\subsection{Supply}\label{app:additional_supply}

\begin{figure}[H]
    \centering
    \caption{Blame and Credit over Topics}
    \label{fig:topic_blamemerit}
    \includegraphics[width=0.6\linewidth]{inputs_paper/current/figures/topics_blamemerit.pdf}
    \caption*{\scriptsize \textit{Notes}: The figure presents the share of blame and credit tweets within each topic. Bars represent 95 percent confidence intervals.}
\end{figure}

\begin{figure}[H]
    \centering
    \caption{Author Correlates of Blame and Credit}
    \label{fig:coefs_author}
    \includegraphics[width=0.5\linewidth]{inputs_paper/current/figures/coefs_author.pdf}
    \vspace{0.3cm}
    \caption*{\scriptsize \textit{Notes}: The figure present the estimates from two regressions at the politician level of blame and credit over the listed variables. Bars represent 95 percent confidence intervals computed with robust standard errors.}
\end{figure}

\begin{figure}[H]
    \centering
    \caption{Causal Rhetoric and Disagreement among Targeted Tweets}
    \label{fig:sentiment_divide_target_pooled}
    \includegraphics[width=0.5\linewidth]{inputs_paper/current/figures/sentiment_divide_targeted_pooled.pdf}
    \caption*{\scriptsize \textit{Notes}: The figure reports evaluative disagreement among targeted tweets, measured as the difference in average standardized sentiment between tweets posted by members of the ruling party and tweets posted by members of the opposition. The sample is restricted to tweets that can be identified as targeting either Democrats or Republicans. Higher values indicate a larger divergence in how politicians on opposite sides of political control evaluate the political environment. We compute this measure separately for all targeted tweets, targeted tweets classified as blame or credit, and targeted tweets classified as neither blame nor credit. Bars represent 95 percent confidence intervals.\par}
\end{figure}

\begin{figure}[H]
    \centering
    \caption{Blame and Credit over Time: Adjusting for Politician and Topic Composition}
    \label{fig:decomposition}
    \begin{subfigure}{0.49\textwidth}
        \caption{Blame}
        \label{fig:decomposition_blame}
        \includegraphics[width=\linewidth]{inputs_paper/current/figures/time_decomposition_blame.pdf}
    \end{subfigure}\hfill
    \begin{subfigure}{0.49\textwidth}
        \caption{Credit}
        \label{fig:decomposition_merit}
        \includegraphics[width=\linewidth]{inputs_paper/current/figures/time_decomposition_merit.pdf}
    \end{subfigure}
    \vspace{0.3cm}
    \caption*{\scriptsize \textit{Notes}: The figure reports yearly changes in the share of tweets classified as blame or credit, relative to 2012. Each series comes from a specification that progressively adds the fixed effects indicated in the legend. Politician fixed effects absorb changes in the composition of members of Congress in the sample. Topic fixed effects absorb changes in the issues politicians discuss. Bars represent 95 percent confidence intervals computed with standard errors clustered at the politician level.}
\end{figure}

\begin{figure}[H]
    \centering
    \caption{Blame and Credit over Time: Excluding Senators}
    \label{fig:time_trend_nosenators}
    \includegraphics[width=0.65\linewidth]{inputs_paper/current/figures/time_both_nosenators.pdf}
    \caption*{\scriptsize \textit{Notes}: The figure reports the yearly share of tweets classified as blame and credit after excluding tweets posted by Senators.}
\end{figure}

\begin{figure}[H]
    \centering
    \caption{Comparison with Other Text Measures over Time}
    \label{fig:time_trend_comparison}
    \includegraphics[width=0.6\linewidth]{inputs_paper/current/figures/time_comparison.pdf}
    \caption*{\scriptsize \textit{Notes}: The figure reports yearly averages of causal rhetoric, sentiment, emotionality, and moral language. All measures are standardized.}
\end{figure}

\FloatBarrier
\newpage
\subsection{Political Returns}\label{app:additional_returns}

\begin{table}[htbp]
\caption{List of Underlying County-Level Covariates}
\label{tab:controls}
\begin{footnotesize}
\centering
\begin{adjustbox}{max width=\textwidth}
\begin{tabular}{l}
\toprule
Variable \\ 
\midrule
Population decile \\
Population density \\
Log(County area) \\
Distance from Austin, TX (in miles) \\
Distance from NYC (in miles) \\
Distance from San Francisco (in miles) \\
Distance from Washington, DC (in miles) \\
\% aged 20--24 \\
\% aged 25--29 \\
\% aged 30--34 \\
\% aged 35--39 \\
\% aged 40--44 \\
\% aged 45--49 \\
\% aged 50+ \\
Population growth, 2000--2016 \\
\% white \\
\% black \\
\% native American \\
\% Asian \\
\% Hispanic \\
\% unemployed \\
\% below poverty level \\
\% employed in IT \\
\% employed in construction/real estate \\
\% employed in manufacturing \\
\% with high school degree \\
\% with college education \\
\% watching Fox News \\
\% watching prime time TV \\
\bottomrule
\end{tabular}\end{adjustbox}
\\[0.5em]
\caption*{\scriptsize \textit{Notes}: The table lists the 29 county-level covariates used to construct the principal components that are interacted with blame, credit, and tweet volume in Equations \ref{reg:causal_spec}.\par}
\end{footnotesize}
\end{table}

\begin{table}[H]
\caption{Donations: First Stage}
\label{tab:donations-fs}
\centering
\begin{footnotesize}
\begin{adjustbox}{max width=\textwidth}
\begin{tabular}{l*{2}{c}}
\toprule
& (1) & (2)  \\
\midrule
& Blame x Users & Credit x Users \\
\midrule
Blame x SXSWFollower2007&2.482\sym{***}&-0.000\sym{***}\\
&(0.180)&(0.000)\\
Credit x SXSWFollower2007&-0.000\sym{***}&2.482\sym{***}\\
&(0.000)&(0.180)\\
Blame x SXSWFollower2006&-0.274&0.000\\
&(0.394)&(0.000)\\
Credit x SXSWFollower2006&0.000&-0.274\\
&(0.000)&(0.394)\\
\midrule
Politician x Month FE & \checkmark & \checkmark  \\
Politician x County FE & \checkmark & \checkmark  \\
County x Month FE & \checkmark & \checkmark  \\
Tweet Volume Control & \checkmark & \checkmark  \\
\midrule
Observations &   168,232,932 &   168,232,932 \\
Clusters &         3,108 &         3,108 \\
\bottomrule
\end{tabular}\end{adjustbox}
\\[1em]
\caption*{\scriptsize \textit{Notes}: The table presents first-stage estimates for the preferred 2SLS specification in Equation \ref{reg:causal_spec}. The dependent variable is the interaction between the standardized monthly share of blame tweets and county Twitter users in column (1), and the interaction between the standardized monthly share of credit tweets and county Twitter users in column (2). The main excluded instruments are the corresponding interactions with the log of one plus the number of SXSW followers in the county who joined Twitter in 2007. The table also reports the coefficients on the corresponding interactions with the log of one plus the number of SXSW followers in the county who joined Twitter in 2006. All specifications include politician-by-month, politician-by-county, and county-by-month fixed effects, as well as the interaction between tweet volume and county Twitter users. Standard errors in parentheses are clustered at the county level. *, **, *** denote significance at the 10\%, 5\%, and 1\% levels, respectively.\par}
\end{footnotesize}
\end{table}

\begin{table}[H]
\caption{Donations: OLS Estimates}
\label{tab:donations-ols}
\centering
\begin{footnotesize}
\begin{adjustbox}{max width=\textwidth}
\begin{tabular}{l*{3}{c}}
\toprule
& (1) & (2) & (3) \\
\midrule
& \makecell{Revenue\\from Donations} & \makecell{Number\\of Donors} & \makecell{Average\\per Donor} \\
\midrule
Blame x Users&0.006\sym{***}&0.002\sym{***}&0.004\sym{***}\\
&(0.000)&(0.000)&(0.000)\\
Credit x Users&-0.000&-0.000&-0.000\sym{**}\\
&(0.000)&(0.000)&(0.000)\\
\midrule
Politician x Month FE & \checkmark & \checkmark & \checkmark  \\
Politician x County FE & \checkmark & \checkmark & \checkmark  \\
County x Month FE & \checkmark & \checkmark & \checkmark  \\
Tweet Volume Control & \checkmark & \checkmark & \checkmark  \\
\midrule
Observations &   168,232,932 &   168,232,932 &   168,232,932 \\
Clusters &         3,108 &         3,108 &         3,108 \\
\bottomrule
\end{tabular}\end{adjustbox}
\\[1em]
\caption*{\scriptsize \textit{Notes}: The table presents OLS estimates of Equation \ref{reg:causal_spec}. The outcomes are log-donation revenue in column (1), log-number of donors in column (2), and log-average donation per donor in column (3). The main regressors are the standardized monthly shares of blame and credit tweets posted by the politician, each interacted with the log of one plus the number of Twitter users in the county. All specifications include politician-by-month, politician-by-county, and county-by-month fixed effects, as well as the interaction between tweet volume and county Twitter users. Standard errors in parentheses are clustered at the county level. *, **, *** denote significance at the 10\%, 5\%, and 1\% levels, respectively.\par}
\end{footnotesize}
\end{table}

\begin{table}[H]
\caption{Donations: Reduced Form Estimates}
\label{tab:donations-rf}
\centering
\begin{footnotesize}
\begin{adjustbox}{max width=\textwidth}
\begin{tabular}{l*{3}{c}}
\toprule
& (1) & (2) & (3) \\
\midrule
& \makecell{Revenue\\from Donations} & \makecell{Number\\of Donors} & \makecell{Average\\per Donor} \\
\midrule
Blame x SXSWFollower2007&0.013\sym{***}&0.006\sym{***}&0.008\sym{***}\\
&(0.003)&(0.001)&(0.002)\\
Credit x SXSWFollower2007&0.003\sym{***}&0.001\sym{**}&0.002\sym{***}\\
&(0.001)&(0.000)&(0.001)\\
Blame x SXSWFollower2006&-0.008&-0.003&-0.005\\
&(0.005)&(0.002)&(0.003)\\
Credit x SXSWFollower2006&0.003&0.001&0.002\\
&(0.002)&(0.001)&(0.001)\\
\midrule
Politician x Month FE & \checkmark & \checkmark & \checkmark  \\
Politician x County FE & \checkmark & \checkmark & \checkmark  \\
County x Month FE & \checkmark & \checkmark & \checkmark  \\
Tweet Volume Control & \checkmark & \checkmark & \checkmark  \\
\midrule
Observations &   168,232,932 &   168,232,932 &   168,232,932 \\
Clusters &         3,108 &         3,108 &         3,108 \\
\bottomrule
\end{tabular}\end{adjustbox}
\\[1em]
\caption*{\scriptsize \textit{Notes}: The table presents reduced-form estimates corresponding to Equation \ref{reg:causal_spec}. The outcomes are log-donation revenue in column (1), log-number of donors in column (2), and log-average donation per donor in column (3). The main regressors are the standardized monthly shares of blame and credit tweets posted by the politician, each interacted with the log of one plus the number of SXSW followers in the county who joined Twitter in 2007. The table also reports the coefficients on the corresponding interactions with the log of one plus the number of SXSW followers in the county who joined Twitter in 2006. All specifications include politician-by-month, politician-by-county, and county-by-month fixed effects, as well as the interaction between tweet volume and county Twitter users. Standard errors in parentheses are clustered at the county level. *, **, *** denote significance at the 10\%, 5\%, and 1\% levels, respectively.\par}
\end{footnotesize}
\end{table}

\begin{table}[H]
\caption{Donations: Robustness Checks}
\label{tab:donations-robustness}
\centering
\begin{footnotesize}
\begin{adjustbox}{max width=\textwidth}
\begin{tabular}{l*{9}{c}}
\toprule
& (1) & (2) & (3) & (4) & (5) & (6) & (7) & (8) & (9) \\
\midrule
 & \multicolumn{3}{c}{Revenue from Donations} & \multicolumn{3}{c}{Number of Donors} & \multicolumn{3}{c}{Average per Donor}\\
\cmidrule(lr){2-4} \cmidrule(lr){5-7} \cmidrule(lr){8-10}
Blame x Users&0.006\sym{***}&0.006\sym{***}&0.006\sym{**}&0.003\sym{***}&0.003\sym{***}&0.004\sym{***}&0.003\sym{***}&0.003\sym{***}&0.002\\
&(0.001)&(0.001)&(0.003)&(0.001)&(0.000)&(0.001)&(0.001)&(0.001)&(0.002)\\
Credit x Users&0.002\sym{***}&0.002\sym{***}&0.002\sym{**}&0.001\sym{***}&0.001\sym{***}&0.000&0.001\sym{***}&0.001\sym{***}&0.002\sym{**}\\
&(0.000)&(0.000)&(0.001)&(0.000)&(0.000)&(0.000)&(0.000)&(0.000)&(0.001)\\
Blame x SXSWFollower2006&-0.006&-0.006&-0.003&-0.003&-0.002&-0.001&-0.004&-0.004&-0.002\\
&(0.004)&(0.004)&(0.004)&(0.002)&(0.002)&(0.002)&(0.003)&(0.003)&(0.003)\\
Credit x SXSWFollower2006&0.003&0.002&0.000&0.001&0.001&-0.000&0.002\sym{*}&0.002&0.000\\
&(0.002)&(0.002)&(0.002)&(0.001)&(0.001)&(0.001)&(0.001)&(0.001)&(0.001)\\
\midrule
Politician x Month FE & \checkmark & \checkmark & \checkmark & \checkmark & \checkmark & \checkmark & \checkmark & \checkmark & \checkmark  \\
Politician x County FE & \checkmark & \checkmark & \checkmark & \checkmark & \checkmark & \checkmark & \checkmark & \checkmark & \checkmark  \\
County x Month FE & \checkmark & \checkmark & \checkmark & \checkmark & \checkmark & \checkmark & \checkmark & \checkmark & \checkmark  \\
Tweet Volume Control & \checkmark & \checkmark & \checkmark & \checkmark & \checkmark & \checkmark & \checkmark & \checkmark & \checkmark  \\
Sentiment Control & \checkmark & & & \checkmark & & & \checkmark & & \\
Attack/Praise Control & & \checkmark & & & \checkmark & & & \checkmark & \\
County Covariates Controls & & & \checkmark & & & \checkmark & & & \checkmark \\
\midrule
Observations &   168,232,932 &   168,232,932 &   168,178,803 &   168,232,932 &   168,232,932 &   168,178,803 &   168,232,932 &   168,232,932 &   168,178,803 \\
Clusters &         3,108 &         3,108 &         3,107 &         3,108 &         3,108 &         3,107 &         3,108 &         3,108 &         3,107 \\
F statistic & 38.19 & 38.19 & 20.20 & 38.19 & 38.19 & 20.20 & 38.19 & 38.19 & 20.20 \\
\bottomrule
\end{tabular}\end{adjustbox}
\\[1em]
\caption*{\scriptsize \textit{Notes}: The table presents 2SLS estimates of Equation \ref{reg:causal_spec} under alternative control specifications. Columns (1)--(3) use log-donation revenue as the outcome, columns (4)--(6) use log-number of donors, and columns (7)--(9) use log-average donation per donor. Within each outcome, the first column adds controls for exposure to non-causal negative and positive sentiment, the second column adds controls for exposure to non-causal attack and praise, and the third column allows the effects of blame, credit, and tweet volume to vary with county covariates. The main regressors are the standardized monthly shares of blame and credit tweets posted by the politician, each interacted with the log of one plus the number of Twitter users in the county. The table also reports the coefficients on the corresponding interactions with the log of one plus the number of SXSW followers in the county who joined Twitter in 2006. Interactions with county Twitter users are instrumented using the corresponding interactions with the log of one plus the number of SXSW followers in the county who joined Twitter in 2007. All specifications include politician-by-month, politician-by-county, and county-by-month fixed effects, as well as the interaction between tweet volume and county Twitter users. Standard errors in parentheses are clustered at the county level. *, **, *** denote significance at the 10\%, 5\%, and 1\% levels, respectively.\par}
\end{footnotesize}
\end{table}

\begin{table}[H]
\caption{Donations: Any Donation}
\label{tab:donations-any}
\centering
\begin{footnotesize}
\begin{adjustbox}{max width=\textwidth}
\begin{tabular}{l*{4}{c}}
\toprule
& (1) & (2) & (3) & (4) \\
\midrule
Blame x Users&0.001\sym{***}&0.001\sym{***}&0.001\sym{***}&0.002\sym{**}\\
&(0.000)&(0.000)&(0.000)&(0.001)\\
Credit x Users&-0.000&-0.000&0.000&0.000\\
&(0.000)&(0.000)&(0.000)&(0.000)\\
Blame x SXSWFollower2006&-0.001&-0.001&-0.001&-0.000\\
&(0.001)&(0.001)&(0.001)&(0.001)\\
Credit x SXSWFollower2006&0.000&0.000&0.000&0.000\\
&(0.000)&(0.000)&(0.000)&(0.000)\\
\midrule
Politician x Month FE & \checkmark & \checkmark & \checkmark & \checkmark  \\
Politician x County FE & \checkmark & \checkmark & \checkmark & \checkmark  \\
County x Month FE & \checkmark & \checkmark & \checkmark & \checkmark  \\
Tweet Volume Control & \checkmark & \checkmark & \checkmark & \checkmark  \\
Sentiment Control & & \checkmark & &  \\
Attack/Praise Control & & & \checkmark &  \\
County Covariates Controls & & & & \checkmark  \\
\midrule
Observations &   168,232,932 &   168,232,932 &   168,232,932 &   168,178,803 \\
Clusters &         3,108 &         3,108 &         3,108 &         3,107 \\
F statistic & 63.65 & 38.19 & 38.19 & 20.20 \\
\bottomrule
\end{tabular}\end{adjustbox}
\\[1em]
\caption*{\scriptsize \textit{Notes}: The table presents 2SLS estimates of Equation \ref{reg:causal_spec} using as outcome an indicator equal to one if the politician receives at least one donation from the county in the month. Column (1) reports the baseline specification with the tweet-volume control. Column (2) adds controls for exposure to non-causal negative and positive sentiment. Column (3) adds controls for exposure to non-causal attack and praise. Column (4) allows the effects of blame, credit, and tweet volume to vary with county covariates. The main regressors are the standardized monthly shares of blame and credit tweets posted by the politician, each interacted with the log of one plus the number of Twitter users in the county. The table also reports the coefficients on the corresponding interactions with the log of one plus the number of SXSW followers in the county who joined Twitter in 2006. Interactions with county Twitter users are instrumented using the corresponding interactions with the log of one plus the number of SXSW followers in the county who joined Twitter in 2007. All specifications include politician-by-month, politician-by-county, and county-by-month fixed effects, as well as the interaction between tweet volume and county Twitter users. Standard errors in parentheses are clustered at the county level. *, **, *** denote significance at the 10\%, 5\%, and 1\% levels, respectively.\par}
\end{footnotesize}
\end{table}

\begin{table}[H]
\caption{Balance in Respondents' Observable Characteristics}
\label{tab:survey-balance-demographics}
\centering
\begin{footnotesize}
\begin{adjustbox}{max width=\textwidth}
\begin{tabular}{l*{5}{c}}
\toprule
& (1) & (2) & (3) & (4) & (5) \\
\midrule
& Female & Age & White & College & Employed \\
\midrule
 & \multicolumn{5}{c}{(a) Blame}\\
\cmidrule(lr){2-6}
Post-Blame&-0.008&-0.200&-0.011\sym{*}&-0.005&-0.006\\
&(0.008)&(0.255)&(0.007)&(0.008)&(0.008)\\
\midrule
District x Month FE  & \checkmark & \checkmark & \checkmark & \checkmark & \checkmark \\
Date FE  & \checkmark & \checkmark & \checkmark & \checkmark & \checkmark \\
Partisan Alignment FE  & \checkmark & \checkmark & \checkmark & \checkmark & \checkmark \\
\midrule
Observations &        20,154 &        20,154 &        20,154 &        20,154 &        20,154 \\
Clusters &           281 &           281 &           281 &           281 &           281 \\
\midrule
 & \multicolumn{5}{c}{(b) Credit} \\ 
\cmidrule(lr){2-6}
Post-Credit&-0.002&-0.305&-0.003&-0.001&0.008\\
&(0.006)&(0.225)&(0.005)&(0.006)&(0.007)\\
\midrule
District x Month FE  & \checkmark & \checkmark & \checkmark & \checkmark & \checkmark \\
Date FE  & \checkmark & \checkmark & \checkmark & \checkmark & \checkmark \\
Partisan Alignment FE  & \checkmark & \checkmark & \checkmark & \checkmark & \checkmark \\
\midrule
Observations &        29,109 &        29,109 &        29,109 &        29,109 &        29,109 \\
Clusters &           325 &           325 &           325 &           325 &           325 \\
\bottomrule
\end{tabular}\end{adjustbox}
\\[1em]
\caption*{\scriptsize \textit{Notes}: The table presents balance checks for respondents' observable characteristics around rhetoric-only days. Each coefficient comes from a separate regression of the column variable on the post-period indicator. Panel (a) reports estimates around blame-only days, and Panel (b) reports estimates around credit-only days. The treatment indicator equals one for respondents interviewed in the seven days after a rhetoric-only day. The outcomes are an indicator for being female, age, an indicator for being white, an indicator for having a college degree, and an indicator for being employed. All regressions include district-by-month, date, and partisan alignment fixed effects. Standard errors in parentheses are clustered at the congressional-district level. *, **, *** denote significance at the 10\%, 5\%, and 1\% levels, respectively.\par}
\end{footnotesize}
\end{table}

\begin{table}[H]
\caption{Balance in Respondent--Politician Partisan Alignment}
\label{tab:survey-balance-ideology}
\centering
\begin{footnotesize}
\begin{adjustbox}{max width=\textwidth}
\begin{tabular}{l*{8}{c}}
\toprule
& (1) & (2) & (3) & (4) & (5) & (6) & (7) & (8) \\
\midrule
& Dem-Dem & Rep-Dem & Ind-Dem & Else-Dem & Dem-Rep & Rep-Rep & Ind-Rep & Else-Rep \\
\midrule
 & \multicolumn{8}{c}{(a) Blame}\\
\cmidrule(lr){2-9}
Post-Blame&-0.010&0.006&0.005&-0.000&-0.003&0.006&-0.004&0.000\\
&(0.007)&(0.007)&(0.005)&(0.003)&(0.004)&(0.004)&(0.003)&(0.002)\\
\midrule
District x Month FE  & \checkmark & \checkmark & \checkmark & \checkmark & \checkmark & \checkmark & \checkmark & \checkmark \\
Date FE  & \checkmark & \checkmark & \checkmark & \checkmark & \checkmark & \checkmark & \checkmark & \checkmark \\
Respondent Covariate FE  & \checkmark & \checkmark & \checkmark & \checkmark & \checkmark & \checkmark & \checkmark & \checkmark\\ 
\midrule
Observations &        20,154 &        20,154 &        20,154 &        20,154 &        20,154 &        20,154 &        20,154 &        20,154 \\
Clusters &           281 &           281 &           281 &           281 &           281 &           281 &           281 &           281 \\
\midrule
 & \multicolumn{8}{c}{(b) Credit} \\ 
\cmidrule(lr){2-9}
Post-Credit&-0.000&0.004&-0.007\sym{*}&0.003&-0.004&0.001&0.005&-0.002\\
&(0.005)&(0.004)&(0.004)&(0.002)&(0.004)&(0.005)&(0.004)&(0.002)\\
\midrule
District x Month FE  & \checkmark & \checkmark & \checkmark & \checkmark & \checkmark & \checkmark & \checkmark & \checkmark \\
Date FE  & \checkmark & \checkmark & \checkmark & \checkmark & \checkmark & \checkmark & \checkmark & \checkmark \\
Respondent Covariate FE  & \checkmark & \checkmark & \checkmark & \checkmark & \checkmark & \checkmark & \checkmark & \checkmark\\ 
\midrule
Observations &        29,109 &        29,109 &        29,109 &        29,109 &        29,109 &        29,109 &        29,109 &        29,109 \\
Clusters &           325 &           325 &           325 &           325 &           325 &           325 &           325 &           325 \\
\bottomrule
\end{tabular}\end{adjustbox}
\\[1em]
\caption*{\scriptsize \textit{Notes}: The table presents balance checks for respondent--politician partisan alignment around rhetoric-only days. Each coefficient comes from a separate regression of the column variable on the post-period indicator. Panel (a) reports estimates around blame-only days, and Panel (b) reports estimates around credit-only days. The treatment indicator equals one for respondents interviewed in the seven days after a rhetoric-only day. The outcomes are indicators for the interaction between respondent party identification and the politician's party. The first term in each column label denotes the respondent's party identification; the second term denotes the politician's party. All regressions include district-by-month, date, and respondent covariate fixed effects. Standard errors in parentheses are clustered at the congressional-district level. *, **, *** denote significance at the 10\%, 5\%, and 1\% levels, respectively.\par}
\end{footnotesize}
\end{table}

\begin{figure}[H]
    \centering
    \caption{Rhetoric-Only Days in the Survey Sample}
    \label{fig:survey-rhetoric-only}

    \begin{subfigure}{0.48\linewidth}
        \caption{Frequency of Rhetoric-Only Days}
        \label{fig:survey-rhetoric-only-frequency}
        \includegraphics[width=\linewidth]{inputs_paper/current/figures/nationscape_rhetoric_only_frequency.pdf}
    \end{subfigure}
    \hspace{0.01\textwidth}
    \begin{subfigure}{0.48\linewidth}
        \caption{Politician Coverage}
        \label{fig:survey-rhetoric-only-politicians}
        \includegraphics[width=\linewidth]{inputs_paper/current/figures/nationscape_rhetoric_only_politicians.pdf}
    \end{subfigure}

    \vspace{0.3cm}
    \caption*{\scriptsize \textit{Notes}: The figure reports the frequency and politician coverage of rhetoric-only days in the survey analysis sample. Panel (a) reports the share of politician-days classified as blame-only, credit-only, or either type in the analysis sample. Either includes days classified as blame-only or credit-only. The bars labeled ``All Posting Days'' use all politician-days with at least one tweet as the denominator. The bars labeled ``Days with Blame or Credit'' restrict the denominator to politician-days with at least one tweet classified as blame or credit. Panel (b) reports the share of politicians who experience at least one blame-only day, at least one credit-only day, or at least one rhetoric-only day of either type in the analysis sample.}
\end{figure}

\begin{figure}[H]
    \centering
    \caption{Voting Intentions: Dynamics}
    \label{fig:survey_dynamic}
    \begin{subfigure}{0.48\linewidth}
        \caption{Blame}
        \label{fig:survey_dynamic_blame}
        \includegraphics[width=\linewidth]{inputs_paper/current/figures/nationscape_event_3d_blame.pdf}
    \end{subfigure}
    \hspace{0.01\textwidth}
    \begin{subfigure}{0.48\linewidth}
        \caption{Credit}
        \label{fig:survey_dynamic_credit}
        \includegraphics[width=\linewidth]{inputs_paper/current/figures/nationscape_event_3d_credit.pdf}
    \end{subfigure}

    \vspace{0.3cm}
    \caption*{\scriptsize \textit{Notes}: The figure presents dynamic estimates around rhetoric-only days. Panel (a) reports estimates around blame-only days, and Panel (b) reports estimates around credit-only days. The omitted category is the rhetoric-only day and the two days immediately before it, corresponding to days $-2$ to $0$. The remaining coefficients are three-day event-time bins before and after the omitted period. The specification includes the fixed effects in Equation \ref{reg:survey}. Observations that fall into more than one event-time bin are excluded. Bars represent 95 percent confidence intervals computed with standard errors clustered at the congressional-district level.}
\end{figure}

\begin{table}[H]
\caption{Voting Intentions: Date Fixed-Effect Robustness}
\label{tab:survey-date-robustness}
\centering
\begin{footnotesize}
\begin{adjustbox}{max width=\textwidth}
\begin{tabular}{l*{6}{c}}
\toprule
& (1) & (2) & (3) & (4) & (5) & (6) \\
\midrule
& \multicolumn{2}{c}{All Respondents} & \multicolumn{4}{c}{Split by Social Media Use} \\
\cmidrule(lr){4-7}
& & & \multicolumn{2}{c}{No Social Media Use} &  \multicolumn{2}{c}{Social Media Use} \\
\midrule
 & \multicolumn{6}{c}{(a) Blame}\\
\cmidrule(lr){2-7}
Post-Blame&0.011\sym{**}&0.009\sym{*}&0.002&0.007&0.015\sym{**}&0.013\sym{**}\\
&(0.005)&(0.005)&(0.011)&(0.011)&(0.006)&(0.006)\\
\midrule
District x Month FE  & \checkmark & \checkmark & \checkmark & \checkmark & \checkmark & \checkmark \\
Partisan Alignment FE  & \checkmark & \checkmark & \checkmark & \checkmark & \checkmark & \checkmark \\
Respondent Covariate FE  & \checkmark & \checkmark & \checkmark & \checkmark & \checkmark & \checkmark\\ 
Date x Politician's Party FE  & \checkmark &  & \checkmark &  & \checkmark &  \\
Date x Respondent's Party FE & & \checkmark & & \checkmark & & \checkmark \\
\midrule
Observations &        20,144 &        20,105 &         5,757 &         5,646 &        14,284 &        14,227 \\
Clusters &           281 &           281 &           275 &           275 &           281 &           281 \\
\midrule
 & \multicolumn{6}{c}{(b) Credit} \\ 
\cmidrule(lr){2-7}
Post-Credit&0.006&0.007&0.003&-0.001&0.005&0.007\\
&(0.004)&(0.004)&(0.010)&(0.009)&(0.005)&(0.005)\\
\midrule
District x Month FE  & \checkmark & \checkmark & \checkmark & \checkmark & \checkmark & \checkmark \\
Partisan Alignment FE  & \checkmark & \checkmark & \checkmark & \checkmark & \checkmark & \checkmark \\
Respondent Covariate FE  & \checkmark & \checkmark & \checkmark & \checkmark & \checkmark & \checkmark\\ 
Date x Politician's Party FE  & \checkmark &  & \checkmark &  & \checkmark &  \\
Date x Respondent's Party FE & & \checkmark & & \checkmark & & \checkmark \\
\midrule
Observations &        29,101 &        29,080 &         8,239 &         8,178 &        20,746 &        20,716 \\
Clusters &           325 &           325 &           322 &           322 &           325 &           325 \\
\bottomrule
\end{tabular}\end{adjustbox}
\\[1em]
\caption*{\scriptsize \textit{Notes}: The table presents estimates of Equation \ref{reg:survey} under alternative date fixed-effect specifications. The outcome is an indicator equal to one if the respondent reports intending to vote for the politician, and zero otherwise. Panel (a) reports estimates for blame-only days, and Panel (b) reports estimates for credit-only days. The treatment indicator equals one for respondents interviewed in the seven days after a blame-only or credit-only day. Columns (1) and (2) use all respondents. Columns (3) and (4) restrict the sample to respondents who do not report having seen or heard political news on social media in the past week. Columns (5) and (6) restrict the sample to respondents who do report having seen or heard political news on social media in the past week. Columns (1), (3), and (5) include date-by-politician-party fixed effects, while columns (2), (4), and (6) include date-by-respondent-party fixed effects. All specifications include district-by-month, partisan alignment, and respondent covariate fixed effects. Standard errors in parentheses are clustered at the congressional-district level. *, **, *** denote significance at the 10\%, 5\%, and 1\% levels, respectively.\par}
\end{footnotesize}
\end{table}

\begin{table}[H]
\caption{Voting Intentions: Tweet-Volume Robustness}
\label{tab:survey-volume-robustness}
\centering
\begin{footnotesize}
\begin{adjustbox}{max width=\textwidth}
\begin{tabular}{l*{6}{c}}
\toprule
& (1) & (2) & (3) & (4) & (5) & (6) \\
\midrule
& \multicolumn{2}{c}{All Respondents} & \multicolumn{4}{c}{Split by Social Media Use} \\
\cmidrule(lr){4-7}
& & & \multicolumn{2}{c}{No Social Media Use} & \multicolumn{2}{c}{Social Media Use} \\
\midrule
 & \multicolumn{6}{c}{(a) Blame}\\
\cmidrule(lr){2-7}
Post-Blame&0.010\sym{**}&0.010\sym{**}&0.004&0.006&0.014\sym{**}&0.014\sym{**}\\
&(0.005)&(0.005)&(0.010)&(0.011)&(0.006)&(0.006)\\
\midrule
District x Month FE  & \checkmark & \checkmark & \checkmark & \checkmark & \checkmark & \checkmark \\
Date FE  & \checkmark & \checkmark & \checkmark & \checkmark & \checkmark & \checkmark \\
Partisan Alignment FE  & \checkmark & \checkmark & \checkmark & \checkmark & \checkmark & \checkmark \\
Respondent Covariate FE  & \checkmark & \checkmark & \checkmark & \checkmark & \checkmark & \checkmark\\ 
Rhetoric-Day Volume FE & \checkmark & & \checkmark & & \checkmark & \\
Window Volume Control & & \checkmark & & \checkmark & & \checkmark \\
\midrule
Observations &        20,154 &        20,154 &         5,781 &         5,781 &        14,300 &        14,300 \\
Clusters &           281 &           281 &           275 &           275 &           281 &           281 \\
\midrule
 & \multicolumn{6}{c}{(b) Credit} \\ 
\cmidrule(lr){2-7}
Post-Credit&0.006&0.006&-0.001&-0.000&0.006&0.006\\
&(0.004)&(0.004)&(0.009)&(0.009)&(0.005)&(0.005)\\
\midrule
District x Month FE  & \checkmark & \checkmark & \checkmark & \checkmark & \checkmark & \checkmark \\
Date FE  & \checkmark & \checkmark & \checkmark & \checkmark & \checkmark & \checkmark \\
Partisan Alignment FE  & \checkmark & \checkmark & \checkmark & \checkmark & \checkmark & \checkmark \\
Respondent Covariate FE  & \checkmark & \checkmark & \checkmark & \checkmark & \checkmark & \checkmark\\ 
Rhetoric-Day Volume FE & \checkmark & & \checkmark & & \checkmark & \\
Window Volume Control & & \checkmark & & \checkmark & & \checkmark \\
\midrule
Observations &        29,109 &        29,109 &         8,257 &         8,257 &        20,756 &        20,756 \\
Clusters &           325 &           325 &           322 &           322 &           325 &           325 \\
\bottomrule
\end{tabular}\end{adjustbox}
\\[1em]
\caption*{\scriptsize \textit{Notes}: The table presents estimates of Equation \ref{reg:survey} under alternative date fixed-effect specifications. The outcome is an indicator equal to one if the respondent reports intending to vote for the politician, and zero otherwise. Panel (a) reports estimates for blame-only days, and Panel (b) reports estimates for credit-only days. The treatment indicator equals one for respondents interviewed in the seven days after a blame-only or credit-only day. Columns (1) and (2) use all respondents. Columns (3) and (4) restrict the sample to respondents who do not report having seen or heard political news on social media in the past week. Columns (5) and (6) restrict the sample to respondents who do report having seen or heard political news on social media in the past week. Columns (1), (3), and (5) absorb fixed effects for raw tweet volume on the rhetoric-only day, while columns (2), (4), and (6) control for thre log of one plus tweet volume across the full comparison window. All specifications include district-by-month, date, partisan alignment, and respondent covariate fixed effects. Standard errors in parentheses are clustered at the congressional-district level. *, **, *** denote significance at the 10\%, 5\%, and 1\% levels, respectively.\par}
\end{footnotesize}
\end{table}

\begin{table}[H]
\caption{Voting Intentions: 99th-Percentile Rhetoric Days}
\label{tab:survey-p99}
\centering
\begin{footnotesize}
\begin{adjustbox}{max width=\textwidth}
\begin{tabular}{l*{3}{c}}
\toprule
& (1) & (2) & (3) \\
\midrule
& \multicolumn{1}{c}{All Respondents} & \multicolumn{2}{c}{Split by Social Media Use} \\
\cmidrule(lr){3-4}
& & No Social Media Use &  Social Media Use \\
\midrule
 & \multicolumn{3}{c}{(a) Blame}\\
\cmidrule(lr){2-4}
Post-Blame&0.009\sym{*}&0.007&0.012\sym{**}\\
&(0.005)&(0.010)&(0.006)\\
\midrule
District x Month FE  & \checkmark & \checkmark & \checkmark \\
Date FE  & \checkmark & \checkmark & \checkmark \\
Partisan Alignment FE  & \checkmark & \checkmark & \checkmark \\
Respondent Covariate FE  & \checkmark & \checkmark & \checkmark\\ 
\midrule
Observations &        21,959 &         6,297 &        15,583 \\
Clusters &           315 &           309 &           315 \\
\midrule
 & \multicolumn{3}{c}{(b) Credit} \\ 
\cmidrule(lr){2-4}
Post-Credit&0.005&-0.000&0.005\\
&(0.004)&(0.009)&(0.005)\\
\midrule
District x Month FE  & \checkmark & \checkmark & \checkmark \\
Date FE  & \checkmark & \checkmark & \checkmark \\
Partisan Alignment FE  & \checkmark & \checkmark & \checkmark \\
Respondent Covariate FE  & \checkmark & \checkmark & \checkmark\\ 
\midrule
Observations &        29,921 &         8,496 &        21,330 \\
Clusters &           343 &           340 &           343 \\
\bottomrule
\end{tabular}\end{adjustbox}
\\[1em]
\caption*{\scriptsize \textit{Notes}: The table presents estimates of Equation \ref{reg:survey} using an alternative definition of rhetoric days based on 99th-percentile blame and credit spike indicators. The outcome is an indicator equal to one if the respondent reports intending to vote for the politician, and zero otherwise. Panel (a) reports estimates for blame-spike days, and Panel (b) reports estimates for credit-spike days. The treatment indicator equals one for respondents interviewed in the seven days after a 99th-percentile blame or credit spike day. Column (1) uses all respondents. Column (2) restricts the sample to respondents who do not report having seen or heard political news on social media in the past week. Column (3) restricts the sample to respondents who do report having seen or heard political news on social media in the past week. All specifications include district-by-month, date, partisan alignment, and respondent covariate fixed effects. Standard errors in parentheses are clustered at the congressional-district level. *, **, *** denote significance at the 10\%, 5\%, and 1\% levels, respectively.}
\end{footnotesize}
\end{table}

\begin{table}[H]
\caption{Negative Rhetoric and Returns to Blame}
\label{tab:negativity}
\centering
\begin{footnotesize}
\renewcommand{\arraystretch}{1.15}
\begin{adjustbox}{max width=\textwidth}
\begin{tabular}{@{}l@{\quad}c@{}p{2em}@{}l@{}c@{}}
\toprule
\multicolumn{2}{c}{(a) Donations} & & \multicolumn{2}{c}{(b) Voting Intentions} \\
\cmidrule{1-2}\cmidrule{4-5}
\makebox[0.37\textwidth][l]{Non-Causal Negative x Users} & \makebox[7em][c]{-0.001\sym{***}} & & & Non-Causal Negative \\
\cline{5-5}
 & (0.000) & & \makebox[\dimexpr.185\textwidth+4em\relax][l]{Post-Non-Causal Negative} & \makebox[\dimexpr.185\textwidth+4em\relax][c]{0.006} \\
Non-Causal Positive x Users & -0.002\sym{***} & & & (0.005) \\
\cline{4-5}
 & (0.000) & & District x Month FE & \checkmark \\
Non-Causal Negative x SXSWFollower2006 & -0.000 & & Date FE & \checkmark \\
 & (0.001) & & Partisan Alignment FE & \checkmark \\
Non-Causal Positive x SXSWFollower2006 & 0.001 & & Respondent Covariate FE & \checkmark \\
\cline{4-5}
 & (0.001) & & Observations &        25,545  \\
\cline{1-2}
Politician x Month FE & \checkmark & & Clusters &           330  \\
\cline{4-5}
Politician x County FE & \checkmark & & & Non-Causal Positive \\
\cline{5-5}
County x Month FE & \checkmark & & Post-Non-Causal Positive & -0.001 \\
Tweet Volume Control & \checkmark & & & (0.004) \\
\cmidrule{1-2}\cmidrule{4-5}
Observations &   168,232,932  & & District x Month FE & \checkmark \\
Clusters &         3,108  & & Date FE & \checkmark \\
F statistic & 63.65  & & Partisan Alignment FE & \checkmark \\
& & & Respondent Covariate FE & \checkmark \\
\cline{4-5}
& & & Observations &        30,287  \\
& & & Clusters &           356  \\
\bottomrule
\end{tabular}
\end{adjustbox}
\\[1em]
\caption*{\scriptsize \textit{Notes}: Panel (a) presents 2SLS estimates of Equation \ref{reg:causal_spec} where the main regressors are the standardized monthly shares of Non-Causal Negative and Non-Causal Positive tweets posted by the politician, each interacted with the log of one plus the number of Twitter users in the county. The outcome is log-donation revenue. The table also reports coefficients on the corresponding interactions with the log of one plus the number of SXSW followers in the county who joined Twitter in 2006; interactions with county Twitter users are instrumented using the corresponding 2007 SXSW-follower interactions. Panel (b) presents estimates of Equation \ref{reg:survey} for Non-Causal Negative-only and Non-Causal Positive-only days. The outcome is an indicator equal to one if the respondent reports intending to vote for the politician, and zero otherwise. The treatment indicator equals one for respondents interviewed in the seven days after the corresponding sentiment-only day. All specifications include the fixed effects and controls indicated in the table. Standard errors in parentheses are clustered at the county level in Panel (a) and at the congressional-district level in Panel (b). *, **, *** denote significance at the 10\%, 5\%, and 1\% levels, respectively.\par}
\end{footnotesize}
\end{table}

\FloatBarrier
\newpage

\section{Platform Changes}\label{app:platform}

A natural concern is that the documented rise in causal rhetoric reflects changes in Twitter's platform design rather than a change in how politicians communicate. Two platform changes are especially relevant during our sample period. On February 10, 2016, Twitter introduced algorithmic feed curation, replacing the strictly chronological ordering of tweets. On November 7, 2017, Twitter expanded the character limit from 140 to 280 characters. We find no evidence that either change generated a discontinuous increase in blame or credit tweets.

We first examine the introduction of algorithmic feed curation. The time-series evidence in \Cref{fig:time_trend} already suggests that blame and credit do not increase noticeably in 2016. We test this more directly using a regression discontinuity design centered on February 10, 2016. The outcome is the daily share of tweets classified as blame or credit, measured in a symmetric 90-day window around the platform change. As shown in \Cref{fig:algorithm_rdd}, there is no evidence of a discontinuity at the cutoff. The estimated jump is $-0.035$ and statistically insignificant ($p=0.529$). This result suggests that the introduction of algorithmic feed curation did not contribute meaningfully to the subsequent rise in causal rhetoric.

We next examine the expansion of the character limit. The relevant concern is that the original 140-character constraint may have disproportionately restricted blame or credit tweets, so that the later expansion mechanically enabled more causal rhetoric. \Cref{fig:chars_limit_hist} plots the distribution of tweet length before November 7, 2017. The figure shows visible bunching in the upper part of the distribution, especially between 100 and 120 characters, indicating that many tweets were close to exhausting the original limit. Crucially, however, the pattern is similar across rhetorical categories. Blame, credit, and none tweets display comparable length distributions. If blame or credit had been especially constrained by the 140-character cap, one would expect stronger bunching near the limit for those categories. Instead, their distributions closely resemble that of none tweets.

We then test for a discontinuity around the character-limit expansion. Using the same regression discontinuity design centered on November 7, 2017, with the daily share of blame or credit tweets as the outcome and a symmetric 90-day window around the reform, \Cref{fig:chars_limit_rdd} shows no evidence of a jump at the cutoff. The estimated discontinuity is $0.056$ and statistically insignificant ($p=0.532$). The rise in causal rhetoric is therefore unlikely to be explained by the move from 140 to 280 characters.

Taken together, the evidence does not support the view that platform design mechanically generated the rise in causal rhetoric. Neither algorithmic feed curation nor the character-limit expansion coincides with a discontinuous increase in blame or credit tweets.

\begin{figure}[H]
    \centering
    \caption{Regression Discontinuity Plot around Algorithmic Feed Introduction}
    \label{fig:algorithm_rdd}
    \includegraphics[width=0.5\linewidth]{inputs_paper/current/figures/algorithm_rdd.pdf}
    \caption*{\scriptsize \textit{Notes}: The figure reports the regression discontinuity plot around February 10, 2016, the date Twitter introduced algorithmic feed curation. The outcome is the daily share of tweets classified as blame or credit. The plot is produced using the \texttt{rdrobust} package of \citet{calonico2017rdrobust}.}
\end{figure}

\begin{figure}[H]
    \centering
    \caption{Distribution of Tweet Length before Character Limit Expansion}
    \label{fig:chars_limit_hist}
    \includegraphics[width=0.5\linewidth]{inputs_paper/current/figures/chars_limit_hist.pdf}
    \caption*{\scriptsize \textit{Notes}: The figure shows the distribution of tweet length, measured in characters, for tweets posted before November 7, 2017, when Twitter expanded the character limit from 140 to 280 characters. The distribution is shown separately by rhetorical category.}
\end{figure}

\begin{figure}[H]
    \centering
    \caption{Regression Discontinuity Plot around Character Limit Expansion}
    \label{fig:chars_limit_rdd}
    \includegraphics[width=0.5\linewidth]{inputs_paper/current/figures/chars_limit_rdd.pdf}
    \caption*{\scriptsize \textit{Notes}: The figure reports the regression discontinuity plot around November 7, 2017, the date Twitter expanded the character limit from 140 to 280 characters. The outcome is the daily share of tweets classified as blame or credit. The plot is produced using the \texttt{rdrobust} package of \citet{calonico2017rdrobust}.}
\end{figure}

\FloatBarrier
\newpage

\section{Trump and Causal Rhetoric}\label{app:trump} 

A natural concern is that the post-2016 rise in causal rhetoric reflects a Trump-style shift in congressional communication. The timing makes this concern natural: the sharp increase in blame begins around Trump's entry into office. If members of Congress absorbed Trump's rhetorical style, the rise in blame and credit could be a by-product of his influence on political language rather than a broader shift in congressional communication.

To assess this possibility, we collect Trump's tweets from the \citeauthor{trumparchive} and classify them using the same blame--credit classifier employed throughout the paper. Because Trump's account was suspended in early 2021, these data end in 2020. We therefore restrict the analysis in this appendix to the period through 2020. This restriction is not central for our purpose, since causal rhetoric had already become substantially more common in congressional communication by then. We compare Trump's rhetorical style with that of members of Congress over the same period.

Trump is not an outlier in the use of causal rhetoric. For each member of Congress, we compute the share of tweets classified as blame and credit over the sample period and compare the resulting distributions with Trump's corresponding shares. As shown in \Cref{fig:trump_dist_blame,fig:trump_dist_credit}, Trump's use of blame is close to the congressional average---slightly above the median but slightly below the mean---while his use of credit is unusually low, falling below the 10th percentile of the congressional distribution. In this sense, Trump's rhetoric is not characterized by unusually intensive use of either blame or credit.

Trump also does not appear to anticipate the rise in causal rhetoric over time. \Cref{fig:trump_time_blame,fig:trump_time_credit} compare the yearly shares of blame and credit tweets for Trump and for members of Congress. For blame, Trump's use evolves similarly to the congressional series and exceeds it only in 2018, by about 5 percentage points. For credit, Trump's use remains consistently below the congressional average throughout. This pattern makes it difficult to attribute the broader rise in congressional causal rhetoric to the diffusion of Trump's own rhetorical style.

What distinguishes Trump's rhetoric is not causal communication, but affective messaging. We repeat the same distributional exercise for non-causal tweets with negative and positive sentiment. \Cref{fig:trump_dist_noncausalneg,fig:trump_dist_noncausalpos} show that Trump is an outlier in the share of non-causal tweets with negative sentiment, lying above the 90th percentile of the congressional distribution. His share of non-causal tweets with positive sentiment is also elevated---above the median and close to the 90th percentile---but less extreme. Thus, Trump is distinctive primarily for relying heavily on affective messaging, especially non-causal negative rhetoric.

Overall, the evidence does not support the interpretation that members of Congress simply absorbed a Trump-style form of causal rhetoric after 2016. Trump is not unusually intensive in blame or credit. If anything, his distinctive rhetorical feature is a comparatively greater reliance on non-causal affective messaging, especially negative sentiment.

\begin{figure}[H]
    \centering
    \caption{Trump and Causal Rhetoric}
    \label{fig:trump_dist_causal}
    \begin{subfigure}{0.49\textwidth}
        \caption{Blame}
        \label{fig:trump_dist_blame}
        \includegraphics[width=\linewidth]{inputs_paper/current/figures/trump_all_synt_blame.pdf}
    \end{subfigure}\hfill
    \begin{subfigure}{0.49\textwidth}
        \caption{Credit}
        \label{fig:trump_dist_credit}
        \includegraphics[width=\linewidth]{inputs_paper/current/figures/trump_all_synt_merit.pdf}
    \end{subfigure}
    \vspace{0.3cm}
    \caption*{\scriptsize \textit{Notes}: Each panel shows the distribution, across members of Congress, of the share of tweets classified as blame or credit over the 2012--2020 period. The solid black line marks the congressional mean, the dashed black lines mark the 10th, 50th, and 90th percentiles, and the solid red line marks Trump's corresponding share.}
\end{figure}

\begin{figure}[H]
    \centering
    \caption{Trump and Causal Rhetoric over Time}
    \label{fig:trump_time}
    \begin{subfigure}{0.49\textwidth}
        \caption{Blame}
        \label{fig:trump_time_blame}
        \includegraphics[width=\linewidth]{inputs_paper/current/figures/trump_time_blame.pdf}
    \end{subfigure}\hfill
    \begin{subfigure}{0.49\textwidth}
        \caption{Credit}
        \label{fig:trump_time_credit}
        \includegraphics[width=\linewidth]{inputs_paper/current/figures/trump_time_merit.pdf}
    \end{subfigure}
    \vspace{0.3cm}
    \caption*{\scriptsize \textit{Notes}: Each panel shows the yearly share of tweets classified as blame or credit for members of Congress and for Trump over the 2012--2020 period. The solid black line reports the congressional average and the dashed red line reports Trump's corresponding share. Shaded areas represent 95 percent confidence intervals.}
\end{figure}

\begin{figure}[H]
    \centering
    \caption{Trump and Non-Causal Affective Messaging}
    \label{fig:trump_dist_noncausal}
    \begin{subfigure}{0.49\textwidth}
        \caption{Negative Sentiment}
        \label{fig:trump_dist_noncausalneg}
        \includegraphics[width=\linewidth]{inputs_paper/current/figures/trump_all_noncausal_rob_neg.pdf}
    \end{subfigure}\hfill
    \begin{subfigure}{0.49\textwidth}
        \caption{Positive Sentiment}
        \label{fig:trump_dist_noncausalpos}
        \includegraphics[width=\linewidth]{inputs_paper/current/figures/trump_all_noncausal_rob_pos.pdf}
    \end{subfigure}
    \vspace{0.3cm}
    \caption*{\scriptsize \textit{Notes}: Each panel shows the distribution, across members of Congress, of the share of tweets that are non-causal and negative or positive in sentiment over the 2012--2020 period. The solid black line marks the congressional mean, the dashed black lines mark the 10th, 50th, and 90th percentiles, and the solid red line marks Trump's corresponding share.}
\end{figure}

\FloatBarrier
\newpage

\section{Persuasion Rate}\label{app:persuasion}

To interpret the magnitude of our estimates, we calculate persuasion rates following \citet{dellavigna2010persuasion}:
\begin{align}
    f = \frac{y_t - y_c}{r \times (e_t-e_c)} \times \frac{1}{1-y_0}.
\end{align}
In this expression, $y_t - y_c$ is the difference in the share of the population donating between treated and untreated groups, $y_0$ is the share that would donate absent treatment, $e_t-e_c$ is the difference in exposure between treated and untreated groups, and $r$ is a reach parameter.

In our setting, we can write $y_t-y_c = \bar{\beta} \times y_c = \beta \times \overline{\text{Users}} \times y_c$, where $\beta$ is the semi-elasticity of the number of donors with respect to a specific rhetorical style and $\overline{\text{Users}}$ is average county log Twitter adoption, equal to 5.29 in our data. Since the estimated $\beta$ for credit is close to zero and statistically insignificant, we focus on blame. We estimate $\beta = 0.003$. Following the literature, we assume $y_0=y_c$, and we set $y_c=0.16$ following \citet{boken2026returns}.

The term $r \times (e_t-e_c)$ requires additional discussion. While $e_c$ is mechanically equal to zero, $e_t$ represents the share of the average county that is on Twitter, equal to 0.32 in \citet{boken2026returns}. For our purposes, however, the relevant exposure metric is the share of the average county that could have been exposed to the rhetorical style. Because \citet{boken2026returns} focus on virality, they assume that a viral tweet is seen by everyone on Twitter, effectively setting $r=1$. This is the same assumption that leads to the $0.94\%$ persuasion rate reported in the main text, which should be interpreted as a lower bound.

There are reasons to believe that $r$ could be lower. For instance, \citet{wojcieszak2022most} find that 40\% of American Twitter users follow at least one politician. Assuming that users see only the posts of accounts they follow, and that they see all of them, would imply $r=0.4$ and therefore $f=2.36\%$. To be transparent about this sensitivity, \Cref{fig:persuasion} reports how the implied persuasion rate changes as the assumed reach parameter $r$ varies. The implied magnitude remains meaningfully interpretable as $r$ decreases: even at $r=0.1$, the implied persuasion rate is $f \approx 9.45\%$, smaller than the Fox News effect documented in \citet{dellavigna2007fox}.

\begin{figure}[H]
    \centering
    \caption{Blame's Persuasion Rate and Reach}
    \label{fig:persuasion}
    \includegraphics[width=0.6\linewidth]{inputs_paper/current/figures/persuasionrate_blame.pdf}
    \caption*{\scriptsize \textit{Notes}: The figure plots the implied persuasion rate $f$ for different values of the reach parameter $r \in [0.1,1]$, holding fixed the remaining inputs used in the calibration described in this appendix.}
\end{figure}

\FloatBarrier
\newpage

\section{Blame and Amplification}\label{app:virality}

This appendix provides additional details on the relationship between rhetorical style and amplification discussed in \Cref{sec:amplification}. We proceed in two steps. First, we describe the availability of engagement data and compare the corresponding subsample to the full tweet corpus. Second, we show that blame tweets receive more retweets and are disproportionately likely to appear in the upper tail of the retweet distribution, whereas credit tweets show no comparable pattern.

Engagement data are available only for a subsample of the tweets used in the main analysis. Retweet counts are observed through March 2020 and cover approximately half of all tweets in the sample over that period. \Cref{tab:virality-attrition} compares the full dataset with the engagement subsample. At the politician level, tweets with engagement data come from politicians who are, on average, about one year older and 2 percentage points more likely to be Republican. At the tweet level, differences are likewise modest: tweets with engagement data exhibit a slightly higher share of credit tweets, by about 4 percentage points, while the remaining dimensions are closely aligned. Overall, these comparisons suggest that the engagement subsample is broadly representative of the underlying sample, reducing concerns that the results below are driven by selection.

To study how blame and credit relate to engagement, we estimate the following specification:
\begin{align}
    r_{ipdt} = \beta_1 \text{Blame}_{i} + \beta_2 \text{Credit}_{i}
    + \lambda_{p \times d} + \eta_{t \times d} + \varepsilon_{ipdt},
    \label{reg:virality_base}
\end{align}
where $r_{ipdt}$ is the standardized number of retweets received by tweet $i$, posted by politician $p$ on date $d$ and topic $t$; $\text{Blame}_{i}$ and $\text{Credit}_{i}$ are indicators equal to one if tweet $i$ is classified as blame or credit, respectively; $\lambda_{p \times d}$ denotes politician-by-date fixed effects; and $\eta_{t \times d}$ denotes topic-by-date fixed effects. Standard errors are clustered at the politician level. The identifying comparison is therefore within politician-date, comparing tweets posted by the same politician on the same day, while also absorbing date-specific differences across topics. \Cref{fig:virality_static} reports the estimates. Blame tweets receive, on average, about 0.15 standard deviations more retweets than tweets not classified as blame or credit. By contrast, credit tweets do not receive significantly more retweets than tweets not classified as blame or credit.

We next ask whether this relationship varies across the distribution of retweets. To do so, we re-estimate Equation \ref{reg:virality_base} ten times, each time using as the dependent variable an indicator for whether the tweet falls into a given decile of the retweet distribution. Each regression therefore captures how blame and credit are associated with the probability of appearing in different parts of the retweet distribution. \Cref{fig:virality_quantile} shows that blame tweets are significantly less likely to appear in the lower deciles and increasingly more likely to appear in the right tail, especially in the top decile. Credit tweets, by contrast, are somewhat more likely to appear just above the median, but show no comparable tendency to reach the upper tail of the distribution.

\begin{table}[H]
\caption{Differences Between With/Without Engagement Datasets}
\label{tab:virality-attrition}
\centering
\begin{footnotesize}
\begin{adjustbox}{max width=\textwidth}
\begin{tabular}{lcccc}
\toprule
& All Tweets until 03/2020 & Tweets with Retweet Data & Difference & p-value\\
\midrule
Female&0.275 (0.000)&0.276 (0.000)&0.001&0.196\\
Age&57.711 (0.007)&58.665 (0.010)&0.955&0.000\\
Black&0.076 (0.000)&0.074 (0.000)&-0.002&0.000\\
Bachelor&0.329 (0.000)&0.336 (0.000)&0.007&0.000\\
Master or Higher&0.626 (0.000)&0.617 (0.000)&-0.009&0.000\\
Republican&0.431 (0.000)&0.450 (0.000)&0.019&0.000\\
|Nominate Score|&0.434 (0.000)&0.429 (0.000)&-0.005&0.000\\
Share of Blame Tweets&0.144 (0.000)&0.148 (0.000)&0.004&0.000\\
Share of Credit Tweets&0.168 (0.000)&0.206 (0.000)&0.038&0.000\\
Share of None Tweets&0.687 (0.000)&0.646 (0.000)&-0.042&0.000\\
Share of Tweets about Economy&0.150 (0.000)&0.151 (0.000)&0.001&0.064\\
Share of Tweets about Environment&0.051 (0.000)&0.054 (0.000)&0.004&0.000\\
Share of Tweets about Gender&0.070 (0.000)&0.064 (0.000)&-0.006&0.000\\
Share of Tweets about Gun Control&0.032 (0.000)&0.032 (0.000)&-0.000&0.326\\
Share of Tweets about Healthcare&0.118 (0.000)&0.122 (0.000)&0.003&0.000\\
Share of Tweets about Immigration&0.055 (0.000)&0.059 (0.000)&0.003&0.000\\
Share of Tweets about Police&0.027 (0.000)&0.027 (0.000)&0.000&0.042\\
Share of Tweets about Racial Relations&0.064 (0.000)&0.059 (0.000)&-0.005&0.000\\
Share of Tweets about Other Topics&0.433 0.000&0.432 0.000&-0.001&0.184\\
Observations&2,322,957&1,178,861&&\\
\bottomrule
\end{tabular}
\end{adjustbox}
\caption*{\raggedright \scriptsize \textit{Notes}: Means and standard errors in parentheses. The last column reports the p-value from a two-sided t-test of equality of means.\par}
\end{footnotesize}
\end{table}

\begin{figure}[H]
    \centering
    \caption{Blame, Credit, and Retweets}
    \label{fig:virality_static}
    \includegraphics[width=0.5\linewidth]{inputs_paper/current/figures/virality_static.pdf}
    \caption*{\scriptsize \textit{Notes}: The figure presents the estimates of Equation \ref{reg:virality_base}. The outcome is the standardized number of retweets. Bars represent 95 percent confidence intervals computed with standard errors clustered at the politician level.}
\end{figure}

\begin{figure}[H]
    \centering
    \caption{Blame, Credit, and Retweets' Distribution}
    \label{fig:virality_quantile}
    \includegraphics[width=0.6\linewidth]{inputs_paper/current/figures/virality_p_quantile_blame.pdf}
    \caption*{\scriptsize \textit{Notes}: The figure reports estimates from regressions that use, as the dependent variable, an indicator for whether a tweet falls into each decile of the retweet distribution. Bars represent 95 percent confidence intervals computed with standard errors clustered at the politician level.}
\end{figure}

\FloatBarrier
\newpage

\section{Assessing the Economic Grounding of Causal Rhetoric}\label{app:assessing}

This Appendix provides additional details on the construction and classification of the economic-tweet sample used in Section \ref{sec:content}. It describes the sample, the coding protocol, and the implementation of the AI-assisted classification procedure, and reports illustrative examples.

The sample used in Section \ref{sec:content} consists of tweets classified as blame or credit that our topic classification identifies as being about the economy. This is the same topic-classification procedure used throughout the paper and discussed in Section \ref{sec:supply}. Applying that procedure to tweets classified as blame or credit yields 413,836 economic tweets. We do not impose any additional exclusions before classification. All such tweets are submitted to the AI-assisted coding procedure described below. A small number of requests fail because the model does not return a complete output in the required format. In practice, this concerns 379 tweets, corresponding to about 0.09 percent of the economic-tweet sample. These observations are therefore excluded from the analysis.

Each tweet is classified according to the dominant causal claim it conveys, if any. The coding task proceeds in three steps. First, for each tweet we extract the main causal claim. If no causal claim is present, the tweet is coded as \emph{na}. Second, we identify the implied economic mechanism underlying that claim. Third, we classify the claim according to the extent to which it is supported by mainstream macroeconomic reasoning. The classifier assigns one of five raw labels: \emph{standard}, \emph{minor qualified}, \emph{major qualified}, \emph{unsupported}, and \emph{na}. Throughout, we refer to the \emph{standard} category as \emph{supported}. Observations coded as \emph{na} are excluded from the analysis.\footnote{The \emph{na} category accounts for about 14\% of the economic-tweet sample. This is in line with the performance of our classifier. In the validation set, the classifier attains an average precision of about 0.84, implying that roughly one sixth of tweets classified as blame or credit should not in fact belong to those categories. Against this benchmark, the 14\% share of tweets reassigned to \emph{na} under the stricter macroeconomic coding procedure appears broadly consistent.}

We implement the classification procedure using GPT-5 mini in batch mode under a fixed prompt and output schema. We instruct the model to rely on standard mainstream macroeconomic reasoning at the core undergraduate level because the objective of the exercise is not to adjudicate frontier debates in macroeconomics, but to assess whether a claim is supported by the settled core of economic knowledge. Using this benchmark keeps the coding transparent and reproducible, and avoids making classifications hinge on areas where expert disagreement remains substantial. The model is instructed to act as a careful research coder, to follow the codebook exactly, and to evaluate claims using standard mainstream macroeconomic reasoning. The system prompt reads as follows:

{\scriptsize
\begin{verbatim}
You are a careful research coder.
Follow the XML codebook exactly.
Use standard mainstream bachelor-level macroeconomic reasoning.
<codebook>
\end{verbatim}
}
The full codebook is separately reported in \Cref{lst:codebook}. Then, the user prompt asks the model to classify a tweet by passing the tweet identifier, text, and date in structured form.

\Cref{tab:macroexpert_freqs} reports the distribution of labels in the economic-tweet sample, both pooled and separately for blame and credit tweets. \Cref{tab:sample_tweets_macroexpert} reports illustrative examples of tweets assigned to each classification category.

\begin{table}[H]
\caption{Distribution of Labels}
\label{tab:macroexpert_freqs}
\centering
\begin{footnotesize}
\begin{adjustbox}{max width=\textwidth}
\begin{tabular}{lccc}
\toprule
& \multicolumn{1}{c}{Pooled} & \multicolumn{2}{c}{Split} \\
 \cmidrule(lr){3-4}
&  & Blame & Credit\\
\midrule
Supported&9.23&8.54&9.79\\
Minor Qualified&61.76&46.07&74.54\\
Major Qualified&28.12&43.62&15.50\\
Unsupported&0.89&1.77&0.17\\
\bottomrule
\end{tabular}\end{adjustbox}
\\[1em]
{\raggedright \scriptsize \textit{Notes}: The table reports the distribution of macroeconomic-support labels in the economic-tweet sample. The pooled column reports the shares across all tweets that our topic classification identifies as being about the economy and that receive a non-\emph{na} label under the macroeconomic coding procedure. The split columns report the same shares separately for tweets classified as blame and credit.\par}
\end{footnotesize}
\end{table}

\begin{table}[H]\centering
    \def\sym#1{\ifmmode^{#1}\else\(^{#1}\)\fi}
    \caption{Sample Tweets}
    \label{tab:sample_tweets_macroexpert}
    \begin{adjustbox}{max width=\textwidth}
    \scriptsize
    \begin{tabular}{
        >{\raggedright\arraybackslash}p{0.45\linewidth} 
        >{\raggedright\arraybackslash}p{0.1\linewidth} 
        >{\raggedright\arraybackslash}p{0.45\linewidth}
    }
    \toprule
    Tweet & Label & Justification \\
    \hline
    \rowcolor{gray!10}
    I voted for additional direct payments in May and October, and tomorrow I’ll vote again for \$2,000 checks to help those in need. \$600 is not enough, and I’ll keep fighting for more. But that’s no reason to stop a relief bill that will help people who are struggling. \#SignTheBill & Standard & Cash transfers increase disposable income and consumption, easing financial hardship; standard macro effect under ordinary assumptions without major qualifiers.\\
    Congress should pass my bill to get rid of the federal gas tax for the rest of the year. This is one step we can take right now to bring down high gas prices. & Minor Qualified & Recognized supply-side price effect; pass-through to consumer prices is standard but magnitude and timing depend on oil market, state taxes, retailer margins, and demand; temporary cut may have limited impact.\\
    \rowcolor{gray!10}
    Biden's \$6T budget is reckless spending that doubles down on his far-left policies. Americans can not move forward if Biden's big government lowers paychecks, raises taxes \$ harms our economy. Republicans have a way forward. URL & Major Qualified & Claim links big fiscal package to lower pay and harm; mainstream macro sees fiscal policy can affect income, taxes, labor incentives and crowding out, but effects depend on composition, financing, monetary offset, labor market frictions and timing.\\
    Extending the current spending levels is better than a government shutdown, but it is still bad policy because it damages our national security and wastes taxpayer dollars. & Unsupported & Tweet links extension of current spending to damage to national security and waste, but mainstream macro sees neutral/ambiguous macro effect of continuing spending; claims rely on program-specific allocation and micro/defense budget details.\\
    \toprule
    \end{tabular}
    \end{adjustbox}
    \\[0.5em]
    {\raggedright \scriptsize \textit{Notes}: The table reports illustrative examples of economic tweets assigned to each macroeconomic-support category under the AI-assisted coding procedure described in this Appendix. For each tweet, we report the assigned label together with the corresponding justification.\par}
\end{table}
\begin{lstlisting}[
  breaklines=true,
  basicstyle=\ttfamily\scriptsize,
  language={},
  numbers=none,
  frame=none,
  caption={Codebook},
  label={lst:codebook},
  captionpos=t
]
<codebook name="macro_tweet_classification">
 <task>
  Classify the dominant economic causal claim in one political tweet according to how far mainstream bachelor-level macroeconomics would support it under the tweet's baseline interpretation: fully under ordinary assumptions, with only ordinary qualifications, with substantial but still mainstream qualifications, not plausibly under ordinary mainstream conditions, or not at all because no meaningful economic causal claim is present.
 </task>

 <input>
  <item name="tweet_id">Unique identifier of the tweet.</item>
  <item name="tweet_text">Text of the tweet to classify.</item>
  <item name="tweet_date">Date of the tweet.</item>
 </input>

 <claim_rules>
  <rule>Code the dominant economic causal claim in the tweet.</rule>
  <rule>Code causal logic, not rhetoric, morality, persuasion, or political desirability.</rule>
  <rule>Interpret rhetorical language as its most reasonable economic claim, but do not invent missing detail.</rule>
  <definition>A meaningful economic causal claim states or clearly implies that a policy, political, or economic factor affects an economic outcome.</definition>

  <support_scope_rules>
   <rule>Do not assign "na" because the claim is difficult to verify empirically, because the tweet refers to a specific policy or episode, or because the named policy is not discussed in textbooks.</rule>
   <rule>If no meaningful economic causal claim is present, assign "na".</rule>
   <rule>If a meaningful economic causal claim is present, assign one of "standard", "minor_qualified", "major_qualified", or "unsupported".</rule>
  </support_scope_rules>
 </claim_rules>

 <interpretation_rules>
  <definition name="baseline_interpretation">
   The most reasonable ordinary reading of the tweet's dominant causal claim, without strengthening or weakening it through charitable reinterpretation.
  </definition>

  <definition name="ordinary_economic_assumptions">
   Familiar background assumptions that normally accompany causal reasoning in mainstream undergraduate macroeconomics.
  </definition>

  <definition name="implied_economic_mechanism">
   The abstract economic channel connecting the cause and the outcome, stated independently of the tweet's policy-specific, partisan, geographic, or branded language.
  </definition>

  <rule>Translate the causal claim, if any, into the implied economic mechanism before evaluating support.</rule>
  <rule>Evaluate the tweet at the level of the economic mechanism, not at the level of the named policy label.</rule>
  <rule>Use the specific policy name only to identify the implied economic mechanism, not as the object of support.</rule>
  <rule>Do not invent a mechanism that introduces a new causal bridge not reasonably expressed in the tweet.</rule>
  <rule>Do not rescue a claim merely because some logically possible economic story can be invented for it.</rule>
 </interpretation_rules>

 <knowledge_rules>
  <rule>Use standard mainstream bachelor-level macroeconomics as taught in introductory and intermediate undergraduate courses.</rule>
  <rule>Do not rely on niche, heterodox, or highly specialized theories unless they are part of standard undergraduate macroeconomic instruction.</rule>
  <rule>Evaluate the tweet's causal logic at the level of the implied economic mechanism, not at the level of the named policy, politician, bill, slogan, or episode.</rule>
  <rule>Do not use hindsight, ex post outcomes, or later historical developments to evaluate the claim.</rule>
  <rule>Do not fact-check whether the named policy or event actually produced the stated outcome; assess whether mainstream macroeconomics would support the causal logic.</rule>
 </knowledge_rules>

 <temporal_context>
  <rule>The tweet date may be used only to identify contemporaneous context and the likely target of the claim.</rule>
  <constraint>Do not use later events or ex post outcomes to evaluate the claim.</constraint>
 </temporal_context>

 <labels>
  <label name="standard">
   <definition>
    Use "standard" when mainstream bachelor-level macroeconomics would support the tweet's dominant causal claim under ordinary economic assumptions. The claim may still be simplified, but no relevant additional qualification is needed.
   </definition>
   <inclusion_criteria>
    <item>The implied mechanism is recognized in mainstream bachelor-level macroeconomics.</item>
    <item>The claim is supportable under ordinary, familiar assumptions in undergraduate macroeconomics.</item>
    <item>No relevant contextual qualification or scope restriction is needed.</item>
   </inclusion_criteria>
   <exclusion_criteria>
    <item>The claim is supportable only if important contextual conditions or scope restrictions do substantial work.</item>
    <item>No plausible mainstream macroeconomic condition would support the claim.</item>
   </exclusion_criteria>
  </label>

  <label name="minor_qualified">
   <definition>
    Use "minor_qualified" when the tweet's dominant causal claim maps onto a recognized mainstream macroeconomic mechanism and is broadly in line with its standard direction, but support depends on familiar contextual qualifications, scope restrictions, or ordinary assumptions.
   </definition>
   <inclusion_criteria>
    <item>The implied mechanism is recognized in mainstream undergraduate macroeconomics.</item>
    <item>The claim's direction is broadly consistent with the standard prediction.</item>
    <item>Support requires ordinary caveats about magnitude, timing, pass-through, openness, slack, or policy context.</item>
    <item>The qualifications refine the claim but do not substantially rewrite its baseline meaning.</item>
   </inclusion_criteria>
   <exclusion_criteria>
    <item>The claim is supportable under ordinary assumptions without meaningful qualification.</item>
    <item>Support requires major unstated offsets or a reversal of the standard directional effect.</item>
    <item>No ordinary plausible mainstream condition would support the claim.</item>
   </exclusion_criteria>
  </label>

  <label name="major_qualified">
   <definition>
    Use "major_qualified" when the tweet's dominant causal claim can be supported by a recognized mainstream macroeconomic mechanism only if substantial unstated conditions, limiting assumptions, or offsetting forces hold, but these are still not so narrow or contrarian as to make the claim unsupported.
   </definition>
   <inclusion_criteria>
    <item>The implied mechanism is still recognizable within mainstream undergraduate macroeconomics.</item>
    <item>Support requires substantial qualifications that do heavy interpretive work.</item>
    <item>The baseline reading is not the standard prediction, but a mainstream route to support still exists under non-remote conditions.</item>
    <item>The claim may require specific policy composition, monetary offset, special incidence, or other important conditions not stated in the tweet.</item>
   </inclusion_criteria>
   <exclusion_criteria>
    <item>The claim is supportable under ordinary assumptions or only familiar caveats.</item>
    <item>Support would require narrow, unusual, strongly contrarian, or specially engineered assumptions.</item>
    <item>The model would need to invent a new causal bridge not reasonably expressed in the tweet.</item>
   </exclusion_criteria>
  </label>

  <label name="unsupported">
   <definition>
    Use "unsupported" when no ordinary plausible mainstream undergraduate macroeconomic route supports the tweet's dominant causal claim under its baseline interpretation, or when support would require narrow, unusual, strongly contrarian, or specially engineered assumptions.
   </definition>
   <inclusion_criteria>
    <item>No ordinary plausible mainstream macroeconomic condition would support the claimed relationship.</item>
    <item>The claim's direction conflicts with the standard prediction and can be defended only by unusual or remote offsets.</item>
    <item>The implied mechanism is not recognized in mainstream undergraduate macroeconomics.</item>
    <item>Any defense would require rescuing the tweet beyond its ordinary meaning.</item>
   </inclusion_criteria>
  </label>

  <label name="na">
   <definition>
    Use "na" when the tweet contains no meaningful economic causal claim.
   </definition>
   <inclusion_criteria>
    <item>The tweet is descriptive, moral, partisan, rhetorical, or evaluative without a meaningful economic causal claim.</item>
    <item>No dominant economic causal claim can be extracted from the tweet.</item>
    <item>Any implied causal reading would require inventing details not reasonably expressed in the tweet.</item>
   </inclusion_criteria>
   <exclusion_criteria>
    <item>A recognizable economic causal claim is present.</item>
    <item>The tweet can be classified as "standard", "minor_qualified", "major_qualified", or "unsupported".</item>
   </exclusion_criteria>
  </label>
 </labels>

 <classification_principles>
  <rule>The labels do not measure empirical truth; they measure how far mainstream undergraduate macroeconomics would go in supporting the dominant causal claim under its baseline interpretation.</rule>
  <rule>Do not reward a claim merely because some theoretical possibility exists.</rule>
  <rule>Use "standard" when support follows under ordinary assumptions without meaningful qualification.</rule>
  <rule>Use "minor_qualified" when support requires familiar and ordinary qualifications that refine, but do not substantially alter, the tweet's baseline meaning.</rule>
  <rule>Use "major_qualified" when support requires substantial unstated qualifications or offsetting conditions that still remain within mainstream undergraduate macroeconomics and are not remote or contrarian.</rule>
  <rule>Use "unsupported" when no ordinary plausible mainstream macroeconomic condition would support the claim, or when support would require narrow, unusual, strongly contrarian, or specially engineered assumptions.</rule>
  <rule>Use "na" when no meaningful economic causal claim is present.</rule>
 </classification_principles>

 <language_rules>
  <rule>Write all output fields in English only.</rule>
  <rule>Do not use any non-English characters.</rule>
 </language_rules>

 <coding_procedure>
  <step number="1">Determine whether the tweet contains a meaningful economic causal claim. If not, assign "na", leave all non-applicable fields empty, and stop coding.</step>
  <step number="2">Extract the dominant causal claim in one short sentence.</step>
  <step number="3">Identify the implied economic mechanism.</step>
  <step number="4">Ask whether mainstream bachelor-level macroeconomics supports the claim under ordinary assumptions, supports it only with important qualifications, or does not plausibly support it.</step>
  <step number="5">Assign exactly one classification label: "standard", "qualified", "unsupported", or "na".</step>
  <step number="6">Provide a brief justification grounded in mainstream macroeconomic reasoning, emphasizing why support is standard, minor_qualified, major_qualified, or unsupported.</step>
 </coding_procedure>

 <output>
  <format>Return exactly one valid JSON object.</format>
  <requirements>
   <item>Return only the JSON object.</item>
   <item>Use empty strings where a field is not applicable.</item>
  </requirements>
  <fields>
   <field name="tweet_id">Unique identifier of the tweet.</field>
   <field name="causal_claim_extracted">The dominant causal claim in one short sentence. If none, return an empty string. (max 100 characters)</field>
   <field name="mechanism_identified">The implied economic mechanism in one short sentence. If none, return an empty string. (max 100 characters)</field>
   <field name="classification">One of: standard, minor_qualified, major_qualified, unsupported, na.</field>
   <field name="justification">A concise, schematic justification grounded in mainstream macroeconomic reasoning. Use short phrases. (max 250 characters)</field>
  </fields>
 </output>
</codebook>

\end{lstlisting}

\end{document}